\pdfoutput=1
\documentclass[aps,prl,superscriptaddress,twocolumn]{revtex4-1}

\makeatletter
\newcommand\footnoteref[1]{\protected@xdef\@thefnmark{\ref{#1}}\@footnotemark}
\makeatother

\usepackage[dvipsnames]{xcolor}
\usepackage{hyperref}

\usepackage{ulem}
\usepackage{epsfig,array}
\usepackage{verbatim}
\usepackage[T1]{fontenc}
\usepackage{neuralnetwork}
\usepackage{siunitx,booktabs}
\usepackage{titlesec}
\usepackage{amsmath}
\usepackage{appendix}

\usepackage{xr}


\usepackage{amsmath,amssymb}
\usepackage{graphicx}
\usepackage{grffile} 
\usepackage{footmisc}
\usepackage{color,bm,bbold,soul,ulem}
\usepackage{lipsum}
\usepackage{enumitem}
\usepackage{natbib}

\usepackage{xspace} 
\usepackage{nicefrac} 

\usepackage{ulem}

  \newcommand{\be}{\begin{equation}}
  \newcommand{\ee}{\end{equation}}
  \newcommand{\bw}{\begin{widetext}}
  \newcommand{\ew}{\end{widetext}}

  \newcommand{\Eq}[1]{Eq.\,\eqref{#1}}
  \newcommand{\Eqs}[1]{Eqs.\,\eqref{#1}}

  \newcommand{\Fig}[1]{Fig.\,\ref{#1}}

  \newcommand{\sfrac}[1]{\ensuremath{\nicefrac{#1}}} 

  \def\SPE{\ensuremath{E_0}\xspace} 

\newcommand{\veca}{\mathbf{a}}

\begin{document}
\title{Predicting impurity spectral functions using machine learning}
\author{Erica J. Sturm}
\email[]{esturm@bnl.gov}
\thanks{These authors contributed equally to this work}
\affiliation{Condensed Matter Physics and Materials Science Division, Brookhaven National Laboratory, Upton, New York 11973, USA}

\author{Matthew R. Carbone}
\email[]{mrc2215@columbia.edu}
\thanks{These authors contributed equally to this work}
\affiliation{Department of Chemistry, Columbia University, New York, New York 10027, USA}

\author{Deyu Lu}
\affiliation{Center for Functional Nanomaterials, Brookhaven National Laboratory, Upton, New York 11973, USA}
\author{Andreas Weichselbaum}
\affiliation{Condensed Matter Physics and Materials Science Division, Brookhaven National Laboratory, Upton, New York 11973, USA}
\author{Robert M. Konik}
\affiliation{Condensed Matter Physics and Materials Science Division, Brookhaven National Laboratory, Upton, New York 11973, USA}

\date{\today}

\begin{abstract}
                                                                   
The Anderson Impurity Model (AIM) is a canonical model of quantum many-body physics.  Here we investigate whether machine learning models, both neural networks (NN) and kernel ridge regression (KRR), can accurately predict the AIM spectral function in all of its regimes, from empty orbital, to mixed valence, to Kondo. To tackle this question, we construct two large spectral databases containing approximately 410k and 600k spectral functions of the single-channel impurity problem. We show that the NN models can accurately predict the AIM spectral function in all of its regimes, with point-wise mean absolute errors down to 0.003 in normalized units.  We find that the trained NN models outperform models based on KRR and enjoy a speedup on the order of $10^5$ over traditional AIM solvers. The required size of the training set of our model can be significantly reduced using furthest point sampling in the AIM parameter space, which is important for generalizing our method to more complicated multi-channel impurity problems of relevance to predicting the properties of real materials.

\end{abstract}

\maketitle

\emph{Introduction.} Describing the physics of strongly correlated quantum many-body systems in real material systems is a signature challenge.  In weakly correlated systems like simple metals, semiconductors, and band insulators, the physics is single-particle in nature and tools like Landau-Fermi liquid theory work well.  However in materials where correlations are not weak, the single-particle picture is typically insufficient to describe the physics at low energy scales where emergent, completely novel phenomena can arise.  
Such physics is of greatest interest
because they gift correlated materials with exceptional properties ranging over high temperature superconductivity \cite{cuprate1,cuprate2}, colossal magnetoresistance \cite{Ramirez1997}, heavy fermion behavior \cite{SiSteglich}, immense thermopower \cite{Homes,Chikina}, and huge volume collapses \cite{McMahan1998} to name but a few.  

Measuring the response functions of applied weak external stimuli represents a key means to probe the properties of a strongly correlated material.  However, of all the properties of a strongly correlated system, the response functions are the most difficult to ascertain theoretically.  The response functions require not only knowledge of the ground state properties of a correlated material, but detailed knowledge of its excited state structure together with matrix elements of the observables of interest (for example, electric and heat currents).  Many different theoretical approaches exist for resolving this difficult problem.  Here, our motivational focus is one technique that has shown great promise for being able to categorically describe wide classes of correlated materials: dynamical mean field theory (DMFT) \cite{Metzner89,Georges96}. 

\begin{figure}
    \includegraphics[width=\linewidth]{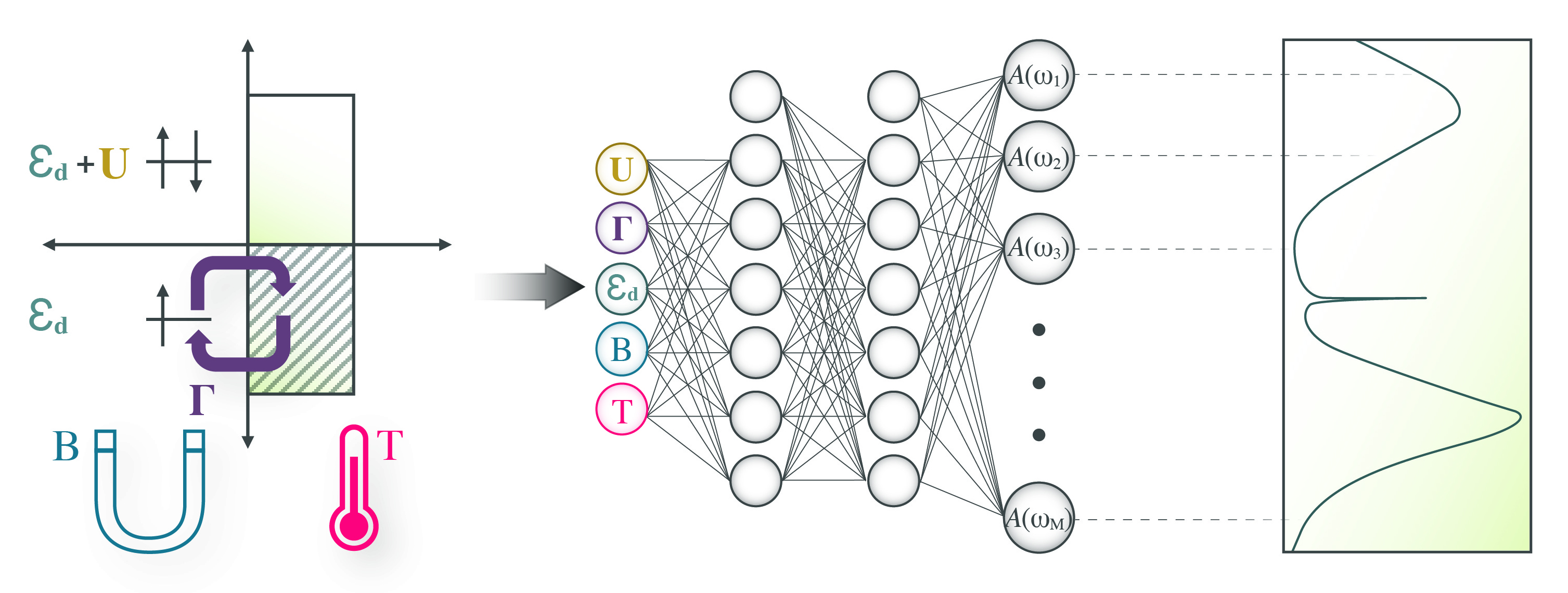}
    \caption{Cartoon of a typical Anderson impurity model
    with a set of physical input parameters (left).
    The physical properties are fed as training data 
    into a neural network (center),
    which then learns to predict
    the spectral function 
    (right).}
    \label{fig:flow-chart}
\end{figure}

DMFT is a Green's function method \cite{Kent348} that sums over an infinite set of Feynman diagrams consistent with the self-energy of the single-particle Green's function being local.  In practice, performing this infinite summation amounts to solving a self-consistent quantum impurity problem.
Besides 
lying at the heart of DMFT,
quantum impurity problems are interesting many-body systems in and of themselves. They also describe magnetic impurities in metallic systems \cite{deHaas34,Costi09}, engineered quantum dots in bosonic or fermionic environments \cite{Leggett87,Cronenwett98,Latta11}, and boundary edge modes in topologically non-trivial systems \cite{Drozdov2014}.  They also generically experience low-energy dynamically generated phenomena that are beyond perturbation theory such as the Kondo effect \cite{Anderson70,Kondo64} and the attendant Abrikosov-Suhl resonance that appears at low temperatures and frequencies in the quantum impurity's spectral response function.

Different techniques are available to find the spectral function of a quantum impurity problem.  Included among those that are numerically exact are continuous-time quantum Monte Carlo simulations \cite{Gull,Aoki14}, the numerical renormalization group (NRG) \cite{Wilson75}, and the density matrix renormalization group \cite{White92}. The number of channels in the impurity problem determines the complexity of computing the spectral function (in the context of DMFT, the number of channels in the effective impurity model matches the number of bands involved in the underlying material).  Single channel impurity models are relatively cheap to solve numerically while multiple channel impurity problems are exponentially more challenging.  For example, five band f-electron materials have associated impurity models requiring petascale computational resources to accurately solve \cite{melnick2020accelerated}.  Furthermore, DMFT embeds quantum impurity model solutions in a self-consistent loop, requiring {\it multiple} solutions for final convergence.

In this light, we ask if machine learning (ML) approaches offer an alternative to expensive many-body simulations of impurity response functions. This question was first posed in Ref.~\onlinecite{Arsenault14}, where kernel ridge regression (KRR) models were trained on a small database of about 5000
spectral functions computed at imaginary frequencies.  A focus of this study was to understand the optimal parameterization of the spectral function for training purposes, finding that a representation in terms of Legendre polynomials worked best. More recently
Ref.~\onlinecite{walker2020neural} used a set of neural networks to train a spectral solver for a quantum impurity connected to a bath of six sites, and each network was trained to predict the spectral function at a single frequency. 

In the work presented herein, we investigate whether an individual model, whether it be KRR or a neural network, can predict the impurity response function in regimes where the relevant energy scales are separated by orders of magnitude and where temperature and magnetic field are also parameters.  To this end, we have constructed large ($\sim10^5$) databases of high fidelity spectral functions in the thermodynamic limit using NRG.  We examine the dependence of these results on the training set size, as generating large training sets for the multi-channel impurity problems, the problem of ultimate interest, is much more computationally intensive.  We use NRG to create our databases as it is an approach that (i) can reliably reach arbitrary, exponentially small dynamically generated energy scales; (ii) computes spectral properties directly on the real frequency axis, and (iii) can work with arbitrary temperatures in an efficient, systematic manner \cite{Bulla08,Anders05,Wb07,Stadler15,Lee17dh}. To the best of our knowledge, such high quality NRG databases do not exist for the AIM model.

\emph{Quantum Impurity Model.} In this work we consider the most elementary of quantum impurity models, the single impurity Anderson model (SIAM) \cite{Anderson61} with a fixed hybridization function.  While we aspire to use ML algorithms
\cite{Rigo20,Hendry19}
to study multi-channel impurity problems, we begin here with the simpler SIAM test environment.  The SIAM Hamiltonian is given by:
\begin{equation}
\label{eqn:SIAM_Ham}
\hat{H} = \hat{H}_\mathrm{imp}
+ \sum_\sigma \int d\varepsilon\, \sqrt{\tfrac{\Gamma(\varepsilon)}{\pi}}
  (\hat{d}_{\sigma}^\dagger \hat{c}_{\varepsilon\sigma} + \mathrm{H.c.})
+ \int d\varepsilon\, \varepsilon \,
\hat{c}_{\varepsilon\sigma}^\dagger \hat{c}_{\varepsilon\sigma}.
\end{equation}
Here, $\hat{H}_\mathrm{imp} = \varepsilon_{d\sigma} \hat{n}_{d\sigma} + U \hat{n}_{d\uparrow} \hat{n}_{d\downarrow}$, where $\hat{d}_\sigma^\dagger$ creates a particle with spin $\sigma\in\{\uparrow,\downarrow\}$ at the impurity $d$-level at energy $\varepsilon_{d\sigma} = \varepsilon_d - \tfrac{\sigma}{2} B$ with $B$ an external magnetic field representing the Zeeman splitting.  Double occupation of the impurity levels pays a Coulombic energy penalty, $U,$ as measured by $U \hat{n}_{d\uparrow} \hat{n}_{d\downarrow}$.
The coupling of the bath of electrons, $\hat{c}_{\varepsilon\sigma}^{(\dagger)},$ to the impurity is described by the hybridization function, $\Gamma(\varepsilon) =\pi\rho_\varepsilon V_\varepsilon^2,$ with $\rho_\varepsilon$ the density of states, and $V_\varepsilon$ the corresponding hopping matrix element at $\varepsilon$. For simplicity in this  work we use a featureless hybridization function, $\Gamma(\varepsilon) = \vartheta(D-|\varepsilon|)\,\Gamma,$ with constant strength $\Gamma=1$ for $|\varepsilon|<D$ and zero elsewhere (we take all energies in units of the half-bandwidth $D=1$, unless specified otherwise, as well as $\hbar=k_B=1$). 

\begin{figure}[t]
    \includegraphics[width=\linewidth]{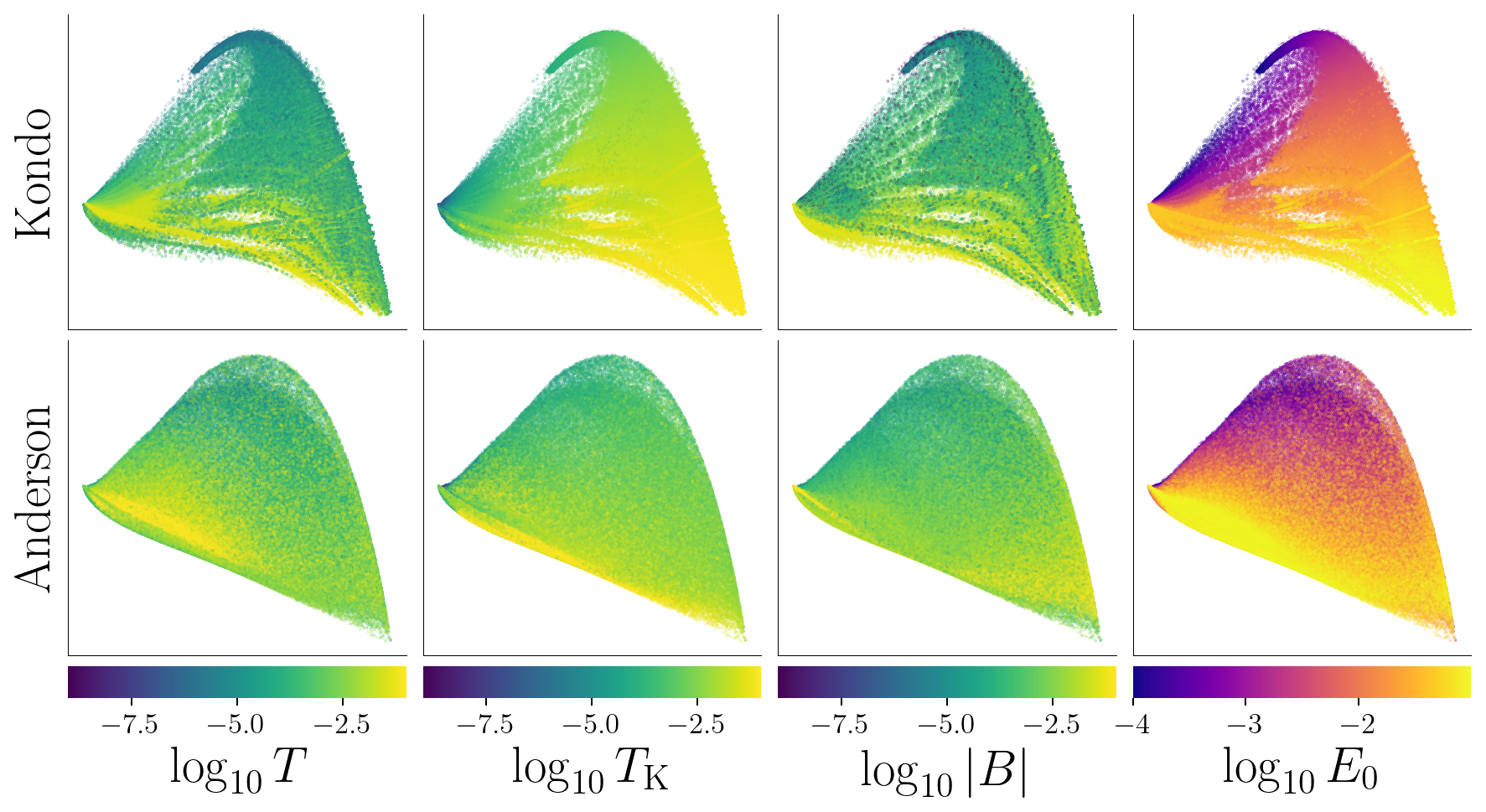}
    \caption{Principal component analysis performed on the Kondo (top) and Anderson (bottom) spectral datasets.
    In each of the four columns the same PCA data are colored in terms of different parameters, left to right: temperature, Kondo temperature, magnetic field strength, and $E_0$.}
\label{fig:PCA}
\end{figure}

\emph{Database Construction.} Inspired by earlier work~\cite{Arsenault14},
we started with impurity parameters, U, $\Gamma$, and $\varepsilon_\mathrm{d}$, which we expanded to include external parameters $B$ and temperature $T$. For a single data point, these inputs will henceforth be written as the ordered set
$x_p \equiv\left(U, \Gamma, \varepsilon_\mathrm{d}, B, T\right)_p.$
We randomly select values for the parameters within predefined physically motivated domains. $x_p$ is mapped onto the scaled target
$y_p(\omega) \equiv \pi \Gamma_p A_p(\omega) \in [0,1]$,
where $A(\omega)$ is the corresponding spectral function
which we compute via the NRG. Two disjoint sets were generated: the ``Anderson"~($\mathcal{D}^\mathrm{A}$) and ``Kondo"~($\mathcal{D}^\mathrm{K}$) sets  
of sizes $|\mathcal{D}^\mathrm{A}| \approx 600\mathrm{k}$ and
$|\mathcal{D}^\mathrm{K}| \approx 410\mathrm{k}.$
Both sets produce related physics and span similar regions of the 5D input hyperspace (see Fig.~\ref{fig:PCA}). As the ML results are similar, we present the Anderson Set results unless otherwise specified. Details on the input parameter generation
including histograms and results on the Kondo Set can
be found in the Appendices~\cite{SM}. 

In practice, spectra were examined between $\omega = \pm 0.8$ to circumvent band edge artifacts, and within that window the spectra are sampled on
a refined mixed linear-logarithmic 
frequency grid $\omega_i$ for $i=1,\ldots,M$ with $M=333$. Finally, all regions of the input hyperspace were sufficiently represented according to the trials' smallest physical energy (SPE) scale, defined as
$E_0 \equiv \max(T, T_\mathrm{K}, |B|),$ where $T_\mathrm{K}$ is the Kondo temperature~\cite{SM}.

Each dataset contains data points 
$(x_p, y_p)$, and is partitioned into disjoint training $(\mathcal{R}^\Omega)$,
cross-validation $(\mathcal{V}^\Omega)$,
and testing $(\mathcal{T}^\Omega)$ sets, with $\Omega \in \{\mathrm{A}, \mathrm{K}\}$ for the Anderson and Kondo datasets, respectively.
We use approximately 2\% of the data for cross-validation, another 2\% for testing,
and the rest for training \footnote{The Anderson and Kondo
data sets are completely disjoint, i.e., they have separate training,
validation and testing sets, and are independently trained and
evaluated. As such we often suppress the index $\Omega$ for brevity.}.
In an effort to evaluate how the data selection method of the minimal training
sets affects the results, we also define two subsets of $\mathcal{R},$ both of size
50k. The first is a randomly down-sampled subset $\mathcal{R}_\mathrm{r} \subset 
\mathcal{R}.$ The second is a furthest-points down-sampled
(FPS) subset
$\mathcal{R}_\mathrm{f} \subset \mathcal{R}$ constructed by first selecting a random
point, and then iteratively sampling the next furthest point in the remainder of the scaled, 5D parameter-space.
Models are hyper parameter-tuned using the validation set, and all results
presented in this work correspond to the testing sets, both of which are consistent
regardless of the model or training set.

In order to visually evaluate the integrity of the generated datasets, we performed a principal component analysis (PCA) in the spectral space, reducing the dimension of each spectrum from
$M$ to $2$. The results are plotted in Fig.~\ref{fig:PCA}, and color-coded with respect to parameters of the SIAM that are most directly relevant for the physical low-energy regime, $T$, $T_\mathrm{K}$, $|B|$, and the derived $E_0$.
Overall, the smooth color gradients observed in the PCA plots suggest that the
physically-relevant input parameters
can be mapped continuously to prominent spectral features.
Additionally, one can identify interesting trends within the physical parameter regimes. In the plots involving $E_0$, the lowest values cluster at the top and towards the left, while the largest $E_0$ scales concentrate at the bottom towards the right/left for the Kondo/Anderson dataset. This suggests that the dynamically generated Kondo peak at low energies and the higher energy side peaks correspond to different spectral features, as already understood from domain knowledge. Despite the overall similarity between the Kondo and Anderson PCA plots, there are subtle differences. For example, in the Kondo set, the input parameters were generated on a grid~\cite{SM}
leading to streaks in the PCA plots while such streaks are absent in the uniformly sampled Anderson set. These PCA plots qualitatively confirm the physical intuition
that a well-defined mapping exists between the input parameters, which
determine the physics of the system, and the spectral functions. This suggests that machine learning algorithms
are well suited to modeling the feature-target
mapping.

\emph{Model Results \& Discussion.} We use the mean absolute error (MAE)
to characterize model performance.
The MAE between a ground truth NRG spectrum
$y_p(\omega)$ and ML-predicted $\hat{y}_p(\omega)$ is defined as
an average over the testing set,
\begin{equation} \label{mae full}
\delta y_p = \frac{1}{M} \sum_{i=1}^M |y_p(\omega_i) - 
    \hat{y}_p(\omega_i)|, \quad
    \overline{\delta y} = \frac{1}{|\mathcal{T}|}
    \sum_{y_p \in \mathcal{T}} \delta y_p,
\end{equation}
as displayed in Table~\ref{tab:results}.

\begin{figure}[b]
    \centering
    \includegraphics[width=\linewidth]{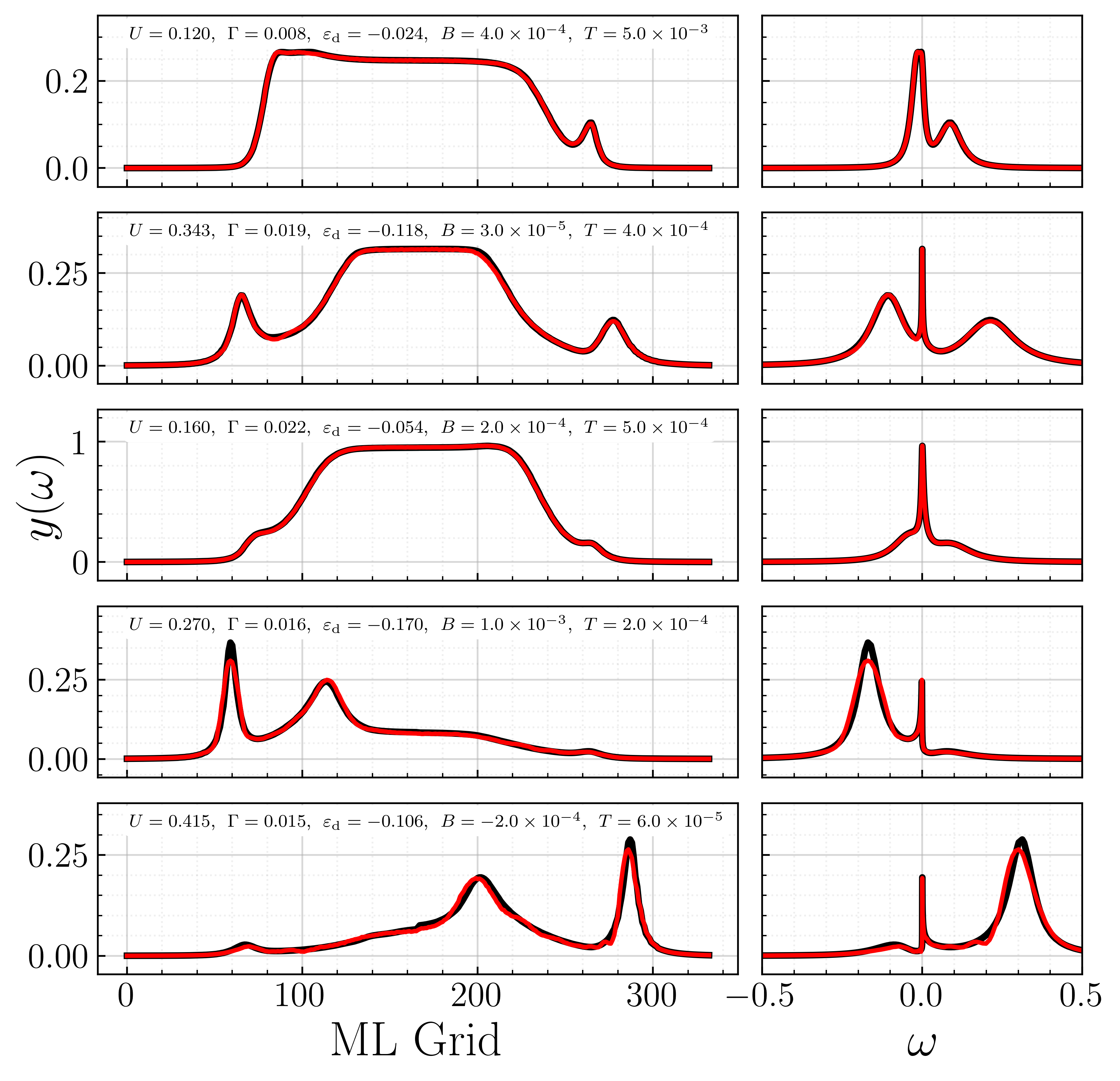}
\caption{ \label{fig:plots from different deciles}
   Representative ground truth (black) and MLP-predicted (red)
   spectral functions from the testing set, $\mathcal{T}^\mathrm{A},$
   from the model trained on $\mathcal{R}^\mathrm{A}.$
   Data correspond to the best example of each of the five pentiles of the data top to bottom, respectively. The system parameters $x_p \equiv\left(U, \Gamma, \varepsilon_\mathrm{d}, B, T\right)_p$ are specified within each panel.
   As the algorithms have no notion of the
   $\omega$-grid on which the spectra are defined, we present the 
   ``ML-grid" (left) which represents the spectral functions, $y_p=\pi\Gamma_p A_p$,
   on a uniformly-spaced grid (emphasizing how the algorithms ``see" the targets).}
\end{figure}

\begin{table*}[t]
\caption{\label{tab:results}%
Summary of average and standard deviations
of mean absolute error [cf. Eq.~\eqref{mae full}]
of all trained models as computed on
test sets, $\mathcal{T}^\Omega$.
The training sets used are shown in parenthesis, e.g., MLP$(\mathcal{R}_\mathrm{f})$ indicates the MLP trained on the down-sampled 50k 
trial $\mathcal{R}_\mathrm{f}$ sets but still evaluated on the
 appropriate $\mathcal{T}$.
All KRR$(\mathcal{R}_\mathrm{r})$
training attempts yielded poor predictions
with $r^2 < 0$. DC-KRR$(\mathcal{R}_\mathrm{F})$ trained on the 
full training set $\mathcal{R}$ 
but in sequentially-sampled subsets ordered by the FPS algorithm as explained in Eq.~\eqref{eqn:R_F}.}
\renewcommand{\arraystretch}{1.5}
\setlength{\tabcolsep}{7pt} 
\begin{tabular}{lc|ccc|ccc}
\hline\hline
$\Omega$ & \textrm{Baseline}&
MLP$(\mathcal{R})$&
MLP$(\mathcal{R}_\mathrm{r})$&
MLP$(\mathcal{R}_\mathrm{f})$&
KRR$(\mathcal{R}_\mathrm{r})$& 
KRR$(\mathcal{R}_\mathrm{f})$&
DC-KRR$(\mathcal{R}_\mathrm{F})$ \\ 
\hline 
A & $0.126 \pm 0.054$ & $0.003 \pm 0.003$ & $0.091 \pm 0.070$ & $0.014 \pm 0.013$ & $0.084 \pm 0.068 $ & $0.021 \pm 0.023$ & $0.017 \pm 0.026$ \\
K & $0.234 \pm 0.076$ & $0.003 \pm 0.002$ & $0.152 \pm 0.141$ & $0.010 \pm 0.009$ & $ 0.152 \pm 0.145 $ &$0.019 \pm 0.021$ & $0.035 \pm 0.033$ \\
\hline\hline
\end{tabular}
\end{table*}

We introduce a measure of the spectral data variation, referred to as the baseline error, as the MAE of the test set against its average spectrum,
$\overline{y}(\omega) \equiv \frac{1}{|\mathcal{T}|} \sum_{y_p \in \mathcal{T}} y_p(\omega)$. 
In a dataset with a high degree of variance, such as $\mathcal{R}^\mathrm{A}$ or
$\mathcal{R}^\mathrm{K}$, we expect the baseline error to be quite large; our
goal is to train  a machine learning model to learn the mapping between the input
and output effectively, and thus significantly outperform this baseline.
We begin our investigation with a multi-layer perceptron (MLP),
a deep learning model capable of capturing highly non-linear
relations in high-dimensional data.
We observe a superb performance from the MLP trained
on $\mathcal{R}$, where the results outperform the baseline by factors of roughly 
40 and 80 on
the Anderson ($\overline{\delta y} = 0.003$) and Kondo ($\overline{\delta y} = 0.003$) testing sets, respectively.

From each pentile of the Anderson test set, the best representative examples of the model predictions are presented in
Fig.~\ref{fig:plots from different deciles}. We first note
that all important spectral features are well-reproduced in
these examples including the peak heights, widths, and locations of the sharp central peak and the broader side peaks.
As expected, models trained using all of
$\mathcal{R}$ present the best results.
However, models trained using 
the FPS subset, $\mathcal{R}_\mathrm{f}$,
which only constitutes about 10\% of each of the full training sets, also
perform surprisingly well, indicating that even a moderately sized training set can result in accurate predictions, if the sampling of the input-parameter space is well-spanned. This is critically important for more complex physical problems where generating training data becomes much more expensive.  By comparison, models trained on a randomly sampled subset, $\mathcal{R}_\mathrm{r},$ of the same size
performs roughly an order of magnitude worse than $\mathcal{R}_\mathrm{f}$, and even barely outperforms the baseline.

We also examine analytical results from the KRR model \cite{scikit-learn, Zhang2013}.
However, this method scales cubically with training set size
$|\mathcal{R}_x|$ with $x\in \{\mathrm{r},\mathrm{f}\},$ as it requires a full matrix inversion, making it intractable to use the full $\mathcal{R}$ at once.
We mitigate this problem in two ways: using down-sampled training sets
of $\mathcal{R}_\mathrm{r}$ or $\mathcal{R}_\mathrm{f}$, and a divide-and-conquer
KRR (DC-KRR) algorithm \cite{Zhang2013, You2018}.
Details regarding KRR and DC-KRR algorithms and hyperparameters can be found in \cite{SM}. 
Models trained using $\mathcal{R}_\mathrm{f}$ perform an order of magnitude better, and have higher coefficients of determination $(r^2)$
than the $\mathcal{R}_\mathrm{r}$ counterparts, in agreement with our earlier
findings for the MLP.

The final KRR models trained on $\mathcal{R}_\mathrm{f}$ are superior to the 
baseline error by factors of roughly 6 and 10 for the Anderson and Kondo sets,
respectively. Interestingly, both the MLP and KRR models trained with the
down-sampled $\mathcal{R}_\mathrm{r}$ achieved nearly identical results, and both are
essentially indistinguishable from the baselines, indicating there is
insufficient information contained in the randomly-downsampled training sets
to train successful models.

In contrast, DC-KRR 
uses the full data set by partitioning $\mathcal{R}$ into $S$ consecutive
(ordered), disjoint subsets $\mathcal{R}_\mathrm{f}^{(s)}$, such that
\begin{equation} \label{eqn:R_F}
    \mathcal{R}_\mathrm{F} = \left(\mathcal{R}_\mathrm{f}^{(1)}, \mathcal{R}_\mathrm{f}^{(2)},
    ..., \mathcal{R}_\mathrm{f}^{(S)} \right), \quad |\mathcal{R}_\mathrm{f}^{(s)}| \approx 50\mathrm{k},
\end{equation}
where the union $\bigcup_s \mathcal{R}_\mathrm{f}^{(s)}$ is equivalent to the full training set $\mathcal{R}$
with respect to the earlier analysis, and it holds that
$\mathcal{R}_\mathrm{f} \equiv \mathcal{R}_\mathrm{f}^{(1)}.$
We then train an independent KRR model for each $\mathcal{R}_\mathrm{f}^{(s)}$ and average the resulting learned parameters.
Because the KRR models performed poorly
with $\mathcal{R}_\mathrm{r}$ we only present the DC-KRR whose subsets were indexed according to the FPS algorithm in Table~\ref{tab:results}. Here, both models exceed the baseline average by a factor of roughly 7.
Despite training on the full data set, the DC-KRR performs comparably to the KRR($\mathcal{R}_\mathrm{f}$) model.

\emph{Conclusion.} In summary we have shown that ML algorithms 
can predict state-of-the-art
Anderson impurity model spectra to overall quantitative accuracy at a speedup of $10^5$ over NRG. We have found that
the use of furthest points sampling can significantly reduce
the required amount of training data to achieve satisfactory
accuracy.
While KRR and DC-KRR are effective for small datasets or large datasets properly divided into chunks of small datasets, our results imply that deep learning algorithms trained on sufficient amount of data are superior for
predicting both the single and many-body features of a SIAM
spectral function.  Future work will expand the physical model to include additional impurity parameters, most importantly
a structured hybridization function and more channels, and thus
examine the viability of a ML algorithm in the context
of a DMFT self-consistent loop.

\begin{acknowledgments}
\emph{Acknowledgments.} EJS and RK were supported by the U.S Department of Energy, Office of Science, Basic Energy Sciences as a part of the Computational Materials Science Program. MRC acknowledges support from the US Department of Energy through the Computational Science Graduate Fellowship (DOE CSGF) under Grant No. DE-FG02-97ER25308. DL was supported by the Center for Functional Nanomaterials, which is a U.S. DOE Office of Science Facility, at Brookhaven National Laboratory under Contract No. DE-SC0012704. AW was supported by the U.S. Department of Energy, Office of Basic Energy Sciences. Resources at the Brookhaven Scientific Data and Computing Center, a component of the Computational Science Initiative were employed.  
We are grateful to Cole Miles, Kipton Barros, and Laura Classen for thoughtful discussions.
\end{acknowledgments}

\appendix
\appendixpage

\setcounter{equation}{0}
\setcounter{figure}{0}
\setcounter{table}{0}
\makeatletter

\renewcommand{\theequation}{A\thesection.\arabic{equation}}
\renewcommand{\thefigure}{A\arabic{figure}}
\renewcommand{\thetable}{A\arabic{table}}

\renewcommand{\appendixname}{}

\renewcommand{\thesection}{A\arabic{subsection}}
\titleformat{\section}[hang]{\large\bfseries}{\thesection}{0.5em}{}
\titleformat{\subsection}[hang]{\bfseries}{\thesection.\thesubsection}{0.5em}{}
\def\p@subsection     {\thesection.}

\begin{appendix}
\appendix\section{A1: Dataset construction}
\label{SI:Dataset_construction}

We use the Numerical Renormalization Group (NRG,
\cite{Wilson75,Bulla08}) to compute the impurity spectral
functions. The NRG obtains the spectral data directly on the
real-frequency axis for arbitrary temperatures where we use
the fdm-NRG approach \cite{Wb07}. We use a typical
discretization parameter 
of $\Lambda=2$, with the discrete data
subsequently smoothened using standard log-Gaussian
broadening schemes. We also compute the local self-energy
$\Sigma^\mathrm{imp}(\omega)$ to improve spectral resolution
of the NRG data \cite{Bulla98}.

Each NRG trial 
takes a set of randomly sampled system
parameters $x_p\equiv\left(U, \Gamma, \varepsilon_\mathrm{d}, B, T\right)_p$, and generates a spectral function
$\pi\Gamma A(\omega)$. Here the factor of $\pi\Gamma$
maintains normalization, such that, e.g., $\pi\Gamma A(0)\le
1$ can be interpreted as transmission probability in
transport measurements. \cite{Meir91, GoldhaberGordon98,
Costi09} The hybridization strength $\Gamma$ is derived
from the hybridization function chosen to be featureless, i.e.
$\Gamma(\omega) = \Gamma \vartheta(D-|\omega|)$,
\cite{Wilson75, Bulla08} where the half-bandwidth $D:=1$
sets the unit of energy throughout, unless specified
otherwise.

Due to the artificial sharp cutoff at the band edge 
caused by our choice of constant hybridization strength,
we only consider the {\it energy window} $\omega = [-0.8,
0.8]\,D$ to avoid artefacts at the band edge. We do not
examine regions, and hence also ignore spectral weight
beyond the band edge. Within the specified window,
we coarse-grain the essentially continuous NRG data to the
same fixed frequency grid to be used in the
machine learning (ML) algorithms. This grid is chosen such that it is
linearly spaced for larger frequencies (67 points for
$\omega\in\left[0.1,0.8\right]$), and logarithmically spaced
for smaller frequencies (99 points for $\omega \in
[ 10^{-5}, 10^{-1} ]$). The same grid is mirrored
for negative frequencies. With the addition of a single point
at $\omega=0$ to bridge the logarithmic grid from positive
to negative we have our final 333 coarse-grain frequency
points. The grid $\omega_i$ is the same for every trial,
which thus maps the five physical parameters to the
normalized spectral function,
\begin{equation}
   x_p\quad \mapsto\quad
   y_{p,i} \equiv \pi\Gamma_p\, A_p(\omega_i).
\end{equation}
The five physical parameters $x_p$ are referred to as the
{\it input features}, and the 333 $y_p$ values as the
{\it output targets} in our ML algorithms. 

The smallest non-zero $\omega$ value in our logarithmic frequency
grid is $10^{-5}$, which is well above the NRG's minimum
set at $\sim10^{-6}$ as determined by our Wilson
chain length of $L=50$ and discretization parameter
$\Lambda=2$. Therefore, to ensure that features are
captured within our frequency grid for ML,
any NRG trials with
smallest physical energy (SPE) \SPE scales
smaller than $10^{-4}$ are removed. The SPE scale 
of each trial is defined by: 
\begin{equation}
  \SPE =\mathrm{max}\left(|B|, T, T_K\right)
\text{ ,}\label{eq:SPE}
\end{equation}
with the Kondo temperature $T_K$ determined
according to Haldane's formula \cite{Haldane78},
\begin{equation}
    T_K = 
    \min\bigl(0.575, \sqrt{\tfrac{U\Gamma}{2}}\bigr)
    \exp\bigl(\tfrac{\pi\varepsilon_d(\varepsilon_d+U)}{2U\Gamma}\bigr)
\text{ .}\label{eq:TKondo}
\end{equation}
By definition of being the smallest physical energy scale, all
physical features in the spectral data are at least as broad as
the SPE.  For example, low-energy Kondo features (described by
$T_K$ which is defined at $B=T=0$) are physically smeared out at the energy scale of $\lvert B \rvert$ or $T$ if these are larger than $T_K$.

\appendix\subsection{A1.1: Physical Parameter Selection}

\begin{figure*}
\centering
\includegraphics[width=\linewidth]{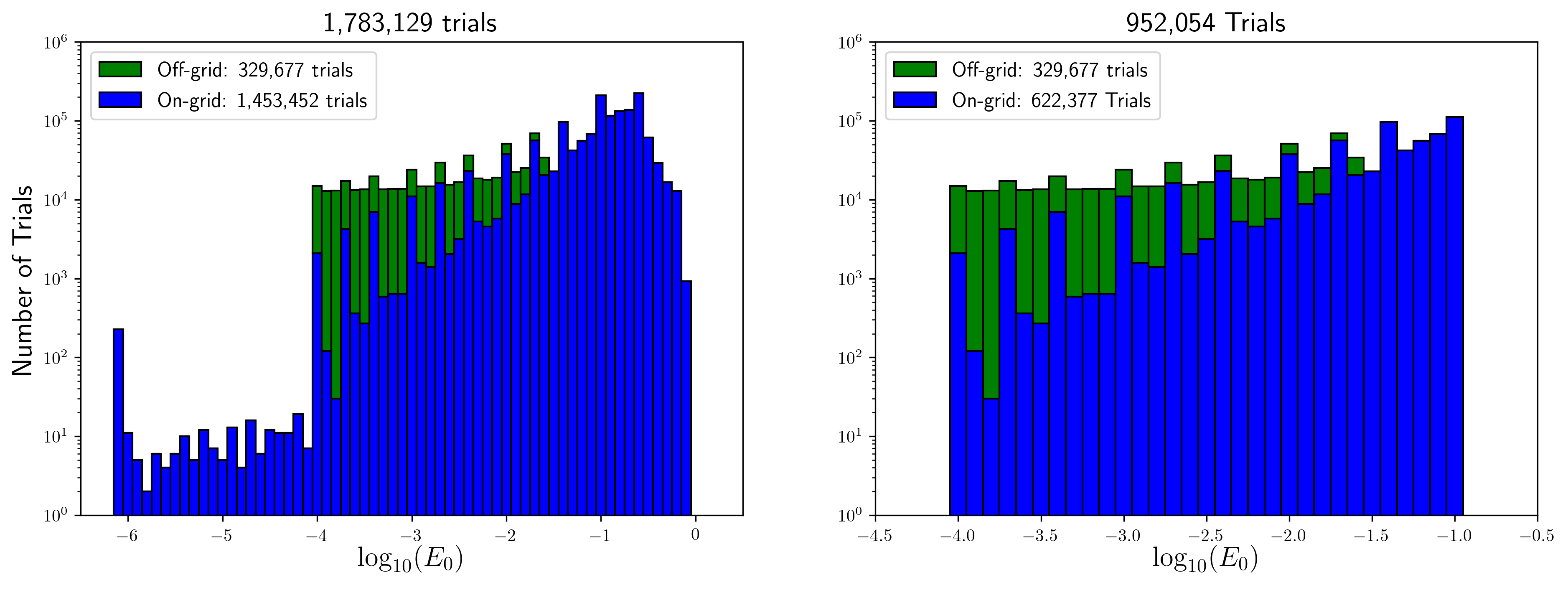}
\caption{SPE histograms for the Kondo data set.  The original {\it on-grid} data (blue) of 1.45 million trials was supplemented by an additional 329k {\it off-grid} trials (green) to ensure a more even \SPE distribution on a log-scale (left panel).  Trials with SPE values outside our range of interest $\SPE \in [10^{-4}, 10^{-1}]$ were discarded, which gave rise to a final SPE distribution (right panel). Bin widths are the same in both panels.} 
\label{fig:ongrid_offgrid_SPE}
\end{figure*}

The five physical parameters for the single impurity Anderson model (SIAM) in this work are collected as an ordered set $x_p \equiv \left(U, \Gamma, \varepsilon_d, B, T\right)_p$.
Inspired by the earlier work of
Arsenault \textit{et. al.} \cite{Arsenault14}, we began with all
five parameters sampled on pre-selected grids (referred to as
`on-grid data' below). This initial set contained just over 1.45
million trials (see blue bars in
Fig.~\ref{fig:ongrid_offgrid_SPE}). However, we eventually found
it favorable to supplement these with 329k additional trials
with random parameter values from preset ranges (`off-grid')
provided that their \SPE values fell within the targeted range
of $[10^{-4}, 10^{-1}]$ (see green bars in
Fig.~\ref{fig:ongrid_offgrid_SPE}). Ultimately we removed any trials where the \SPE did not fall in the desired energy range (see right panel in Fig.~\ref{fig:ongrid_offgrid_SPE}).

\begin{figure*}
\centering
\includegraphics[width=\linewidth]{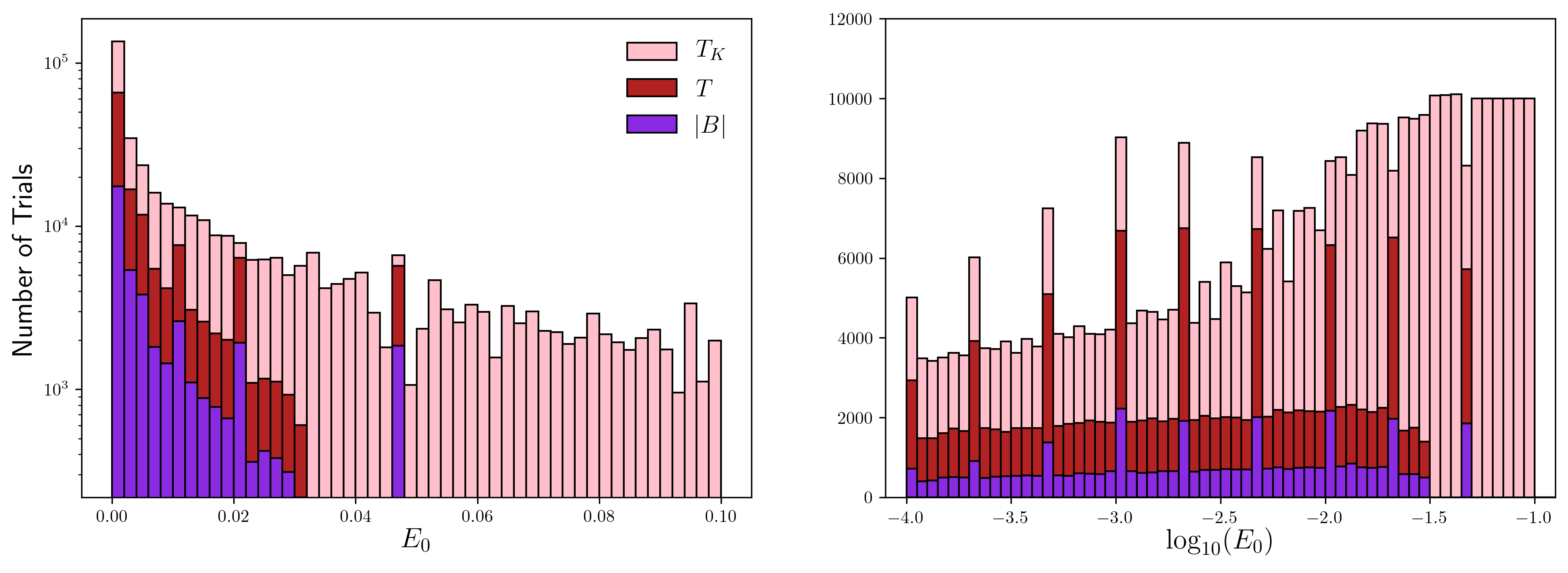}
\caption{SPE histogram of 411,267 data point Kondo set
on linear and log-scale (left and right panel
respectively),
with approximately equal distribution across
trials at lower energies only where 
\SPE is dominated by $|B|$, $T$, or $T_K$. Regularly spaced spikes in right panel are due to the {\it on-grid} portion of the Kondo set.}
\label{fig:K_SPE_cake}
\end{figure*}

We then enforced the additional
requirement on the hybridization strength
$\Gamma$, 
\begin{equation}
    \Gamma > 
    \mathrm{min}\bigl(
      \tfrac{|B|}{5}, 
      \tfrac{U}{40}
    \bigr)
\label{eq:cond:Gamma}
\end{equation}
in order to avoid extremely narrow features in the spectral data comparable to or below the frequency grid spacing chosen for ML. From a physical perspective, this would correspond to an essentially decoupled and hence trivial impurity.
Finally, a randomly selected subset of trials
that satisfied the above requirements were selected for 
a data set of $\sim$~411k trials. We refer to this dataset as the
``Kondo Set'' because
many parameter sets have large Coulomb energies $U$ comparable to or larger than the half bandwidth (i.e. $U\gtrsim1$).
The SPE distribution for the Kondo set is presented in Fig.~\ref{fig:K_SPE_cake}, while the parameter value distribution is shown in black in Fig.~\ref{fig:Kondo_parameter_distribution}.

\begin{figure*}
\centering
\includegraphics[width=\linewidth]{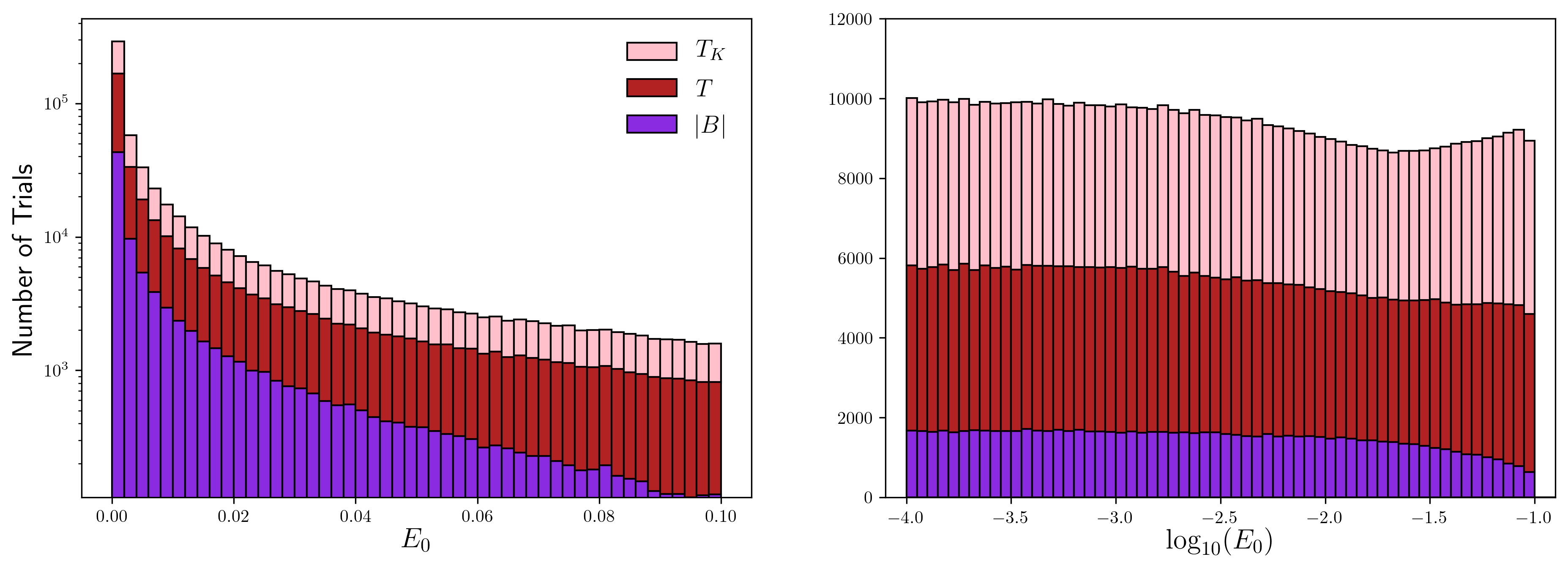}
\caption{SPE histogram of 599,578 data point Anderson set
on linear and log-scale (left and right panel
respectively), with approximately equal distribution across
trials where \SPE is dominated by $|B|$, $T$,
or $T_K$.}
\label{fig:A_SPE_cake}
\end{figure*}

In contrast, the ``Anderson Set'' was designed with
an even SPE distribution in mind, while also keeping
$U<0.5$ right from the start as shown in black in the left-most panels of Fig.~\ref{fig:Anderson_parameter_distribution}. To be specific, the parameters for this data set were sampled
in the following manner: 
(i) randomly choose an SPE value 
with a flat distribution on the log-scale in the range
$\SPE \in [10^{-4},10^{-1}]$
(ii) randomly select which of the three
parameters $|B|$, $T$, or $T_K$ takes that \SPE value, then (iii)
ensure that  the other two parameters 
are smaller than \SPE, using the range $[0,\SPE/10]$.
Both positive and negative values for $B$
are computed in all cases.
With $T_K$ fixed, (iv) the remaining values
for $U$, $\varepsilon_d$, and $\Gamma$ are sampled
based on \Eq{eq:TKondo}. Specifically,
the values for $\Gamma$ and $U$ are assigned
at random on a linear scale below bandwidth,
while still ensuring \Eq{eq:cond:Gamma}.
This fixes the prefactor in \Eq{eq:TKondo}.
By taking the logarithm, we then solve for $\varepsilon_d$.
If no valid solutions exist for $\varepsilon_d$ given the 
choices in $U$ and $\Gamma$, we go back and repeat step (iv).
With this basic algorithm we generated a total of
about 600k trials with the desired flat SPE distribution
as demonstrated in \Fig{fig:A_SPE_cake}. The parameter distribution for the Anderson Set is shown in black in  Fig.~\ref{fig:Anderson_parameter_distribution}.

\appendix\subsection{A1.2: Split selection} 
\label{SI:split_selection} 

The full datasets for the Anderson and Kondo datasets are labeled
$\mathcal{D}^{\Omega}$ with $\Omega \in \{\mathrm{A}, \mathrm{K}\}.$
Each contain about 500k
of data points $(x_p,y_p)$ that are 
split into disjoint training
($\mathcal{R}^{\Omega}$),
validation ($\mathcal{V}^{\Omega}$), and
testing ($\mathcal{T}^{\Omega}$) sets.
Since the Anderson and Kondo sets themselves are disjoint
and also mostly dealt with on an equal footing,
we generally suppress the subscript $\Omega$
for readability unless stated otherwise. 
The two pairs of validation and testing sets
are fixed for the entirety of this work.
The splits were generated as follows:
\begin{enumerate}

\item The testing set, $\mathcal{T}^\Omega,$
    contains roughly 2\% of the total data and is selected by randomly down-sampling $\mathcal{D}^\Omega$. It is evaluated at the end of the pipeline, and represents the most
    unbiased representation of the model performance. All results are evaluated on the testing set, unless explicitly stated otherwise.
    
\item The validation set, $\mathcal{V}^\Omega,$
    also contains
    roughly 2\% of the total data, and is also selected by randomly down-sampling $\mathcal{D}^\Omega.$
    It is used only to tune model hyperparameters.
    
\item The full training set, $\mathcal{R}$ (with $\Omega$ indexing suppressed), contains
96\% of the total data and is used to train the model. The different versions of the training set are explained below.

\begin{itemize} 

\item The randomly sampled training set $\mathcal{R}_\mathrm{r}$ contains 50k data points. It is selected by randomly down-sampling $\mathcal{R}.$

\item The furthest points-sampled (FPS)
    training set $\mathcal{R}_\mathrm{f}$
    contains 50k data points. It is selected via the
    algorithm presented in Section A1.4.

\item For divide-and-conquer DC-KRR only, we 
generate a collection of \emph{ordered} disjoint subsets
$\mathcal{R}_\mathrm{F} \equiv \left(
   \mathcal{R}_\mathrm{f}^{(1)},
   \mathcal{R}_\mathrm{f}^{(2)}, \ldots,
   \mathcal{R}_\mathrm{f}^{(S)}
\right)$, where the set itself
$\mathcal{R} = \bigcup_{s=1}^S \mathcal{R}_\mathrm{f}^{(s)}$
(i.e., the unordered union of the subsets are equivalent to the full training set).
The sequence is ordered in the sense that the
$\mathcal{R}_\mathrm{f}^{(s)}$ are
generated consecutively using the algorithm presented in Section A1.4.
As an example, the sampling of $\mathcal{R}_\mathrm{f}^{(s)}$
excludes samples already included in $\{
   \mathcal{R}_\mathrm{f}^{(1)}, \ldots,
   \mathcal{R}_\mathrm{f}^{(s-1)}\},$
and the first point in $\mathcal{R}_\mathrm{f}^{(s)}$ is furthest-sampled from the
last point in $\mathcal{R}_\mathrm{f}^{(s-1)}.$ Note also that $\mathcal{R}_\mathrm{f}^{(1)} = \mathcal{R}_\mathrm{f}.$
   
\end{itemize}
\end{enumerate}

\appendix\subsection{A1.3: Preprocessing input features (symlog scaling)}
\label{SI:preprocessing}

Anderson and Kondo-type models frequently exhibit
dynamically generated, exponentially small energy scales.
Hence the parameters that enter the SPE in Eq.~\ref{eq:SPE}, 
namely $T_\mathrm{K}$, $B$, and $T$, were sampled on a logarithmic scale that
stretches over several orders of magnitude.
Therefore, prior to model training, one is naturally led
to also apply a (symmetric)
logarithmic (``symlog'') rescaling to the input
features $x_p$.
Feature scaling was necessary for the MLP and the
analytical methods when $\mathcal{R}_\mathrm{R}$
was used, as described elsewhere in this document.

Given that the value for $T_\mathrm{K}$ itself
is not a bare Hamiltonian parameter, we apply
the following logarithmic scaling to
$\Gamma$, $B$, and $T$:
\begin{eqnarray} \label{SI:prescaling eq}
  \Gamma &\to& \log_{10}\,\Gamma \notag \\
  B &\to& \mathrm{symlog}_{10}(B) \equiv
  \mathrm{sgn}(B) \log_{10} |B| \\
  T &\to& \log_{10}\,T  \notag
\text{ ,}\label{eq:symlog}
\end{eqnarray}
where $U$ and $\varepsilon_d$ are left on the linear scale.
Here $\Gamma, T>0$ are always chosen non-zero, yet
possibly exponentially small. Also, we always have
$|B|<1$, and in the case
that $B=0$, the trial's $B$ value is reset to an order of magnitude smaller than the smallest non-negative $B$ value
in the set.

After performing the scaling step in Eq.~\ref{eq:symlog}, the mean ($\mu$) and standard deviation ($\sigma$) are computed for each
of the 5 components $x_p^{(i)} \in x_p$ in the current training set. Then we use
$x_p^{(i)} \rightarrow \frac{x_p^{(i)}-\mu^{(i)}}{\sigma^{(i)}}$
for every trial from the training, validation,
and testing sets.
\appendix\subsection{A1.4: Furthest points sampling algorithm} 
\label{SI:FPS_algorithm}
Here we employ a relatively simple algorithm for sampling data points in an arbitrary dimensional space.
Note that sampling is performed on scaled features
as to treat their notion of distance on equal
footing. The algorithm is defined below:
\begin{enumerate}

\item Select a random point
    $x_p \in \mathcal{R}$ (here, $x_p$ represents the
    5-dimensional input parameter vector, i.e., the feature
    space). It is labeled the \emph{current point},
    and added as first point
    to the set of sampled points $\mathcal{R}_\mathrm{F}$.

\item Find the point that is furthest away from the
    current point that has not yet been selected,
    and add it to the set of sampled points.
    We define {\it distance} by
    $||\cdot||_{L_d}$ with $d=2$.

\item Repeat step 2 until the desired size of $\mathcal{R}_\mathrm{F}$ is reached.

\end{enumerate}

\appendix\subsection{A1.5: Parameter Distributions}
\label{SI:Parameter_Dist}

\begin{figure*}
\centering
\includegraphics[width=0.33\linewidth]{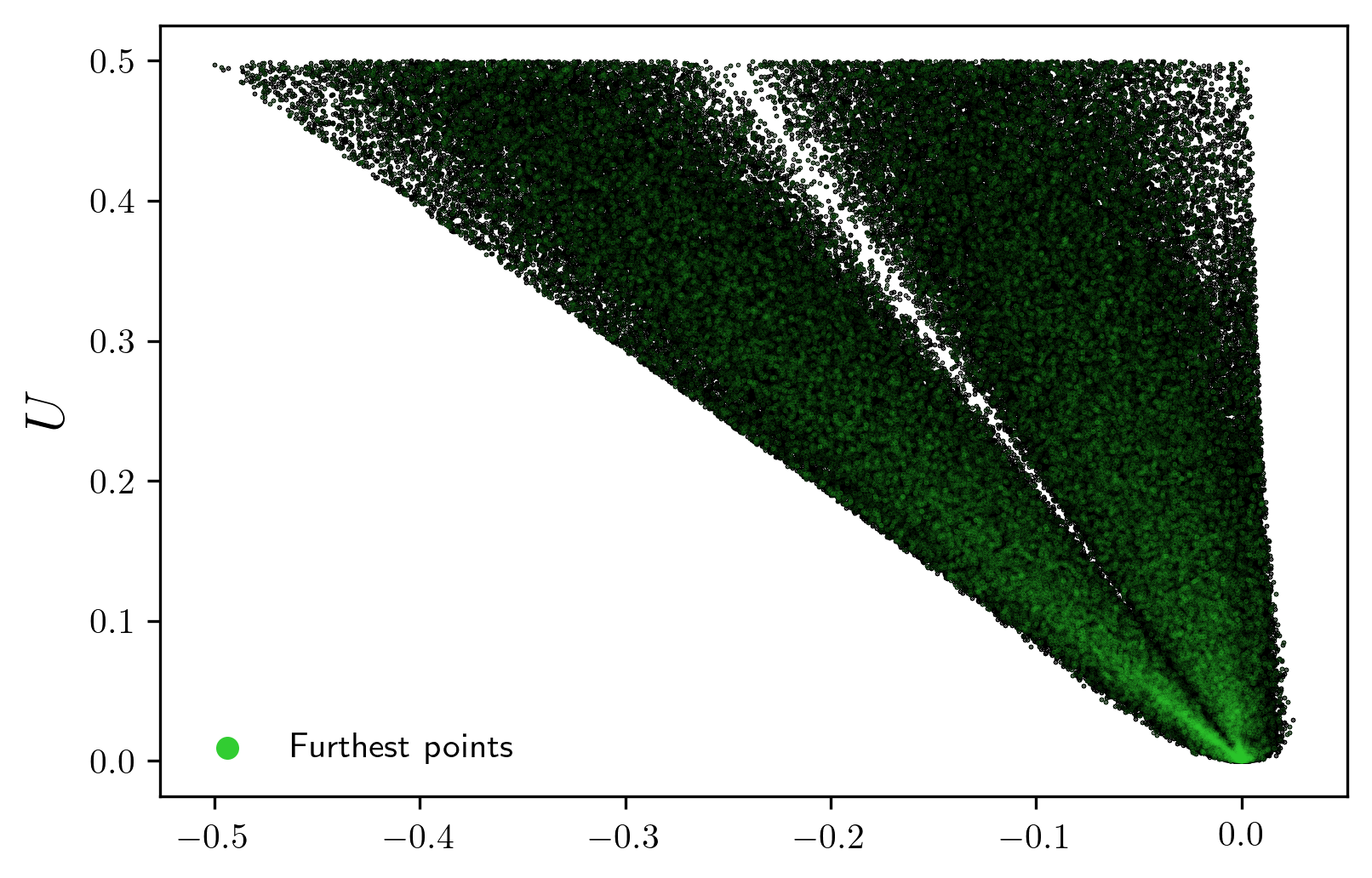}%
\includegraphics[width=0.33\linewidth]{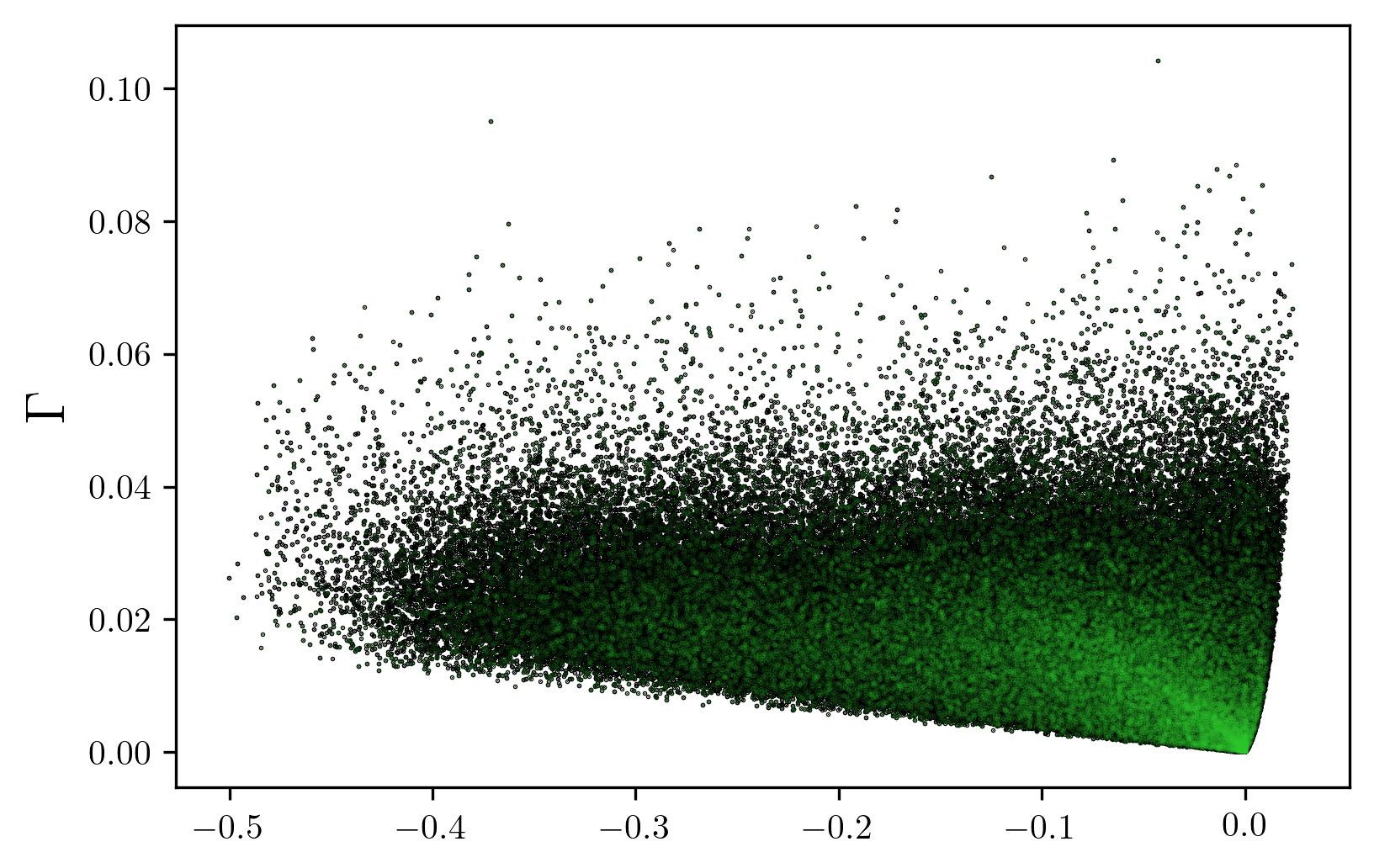}
\includegraphics[width=0.33\linewidth]{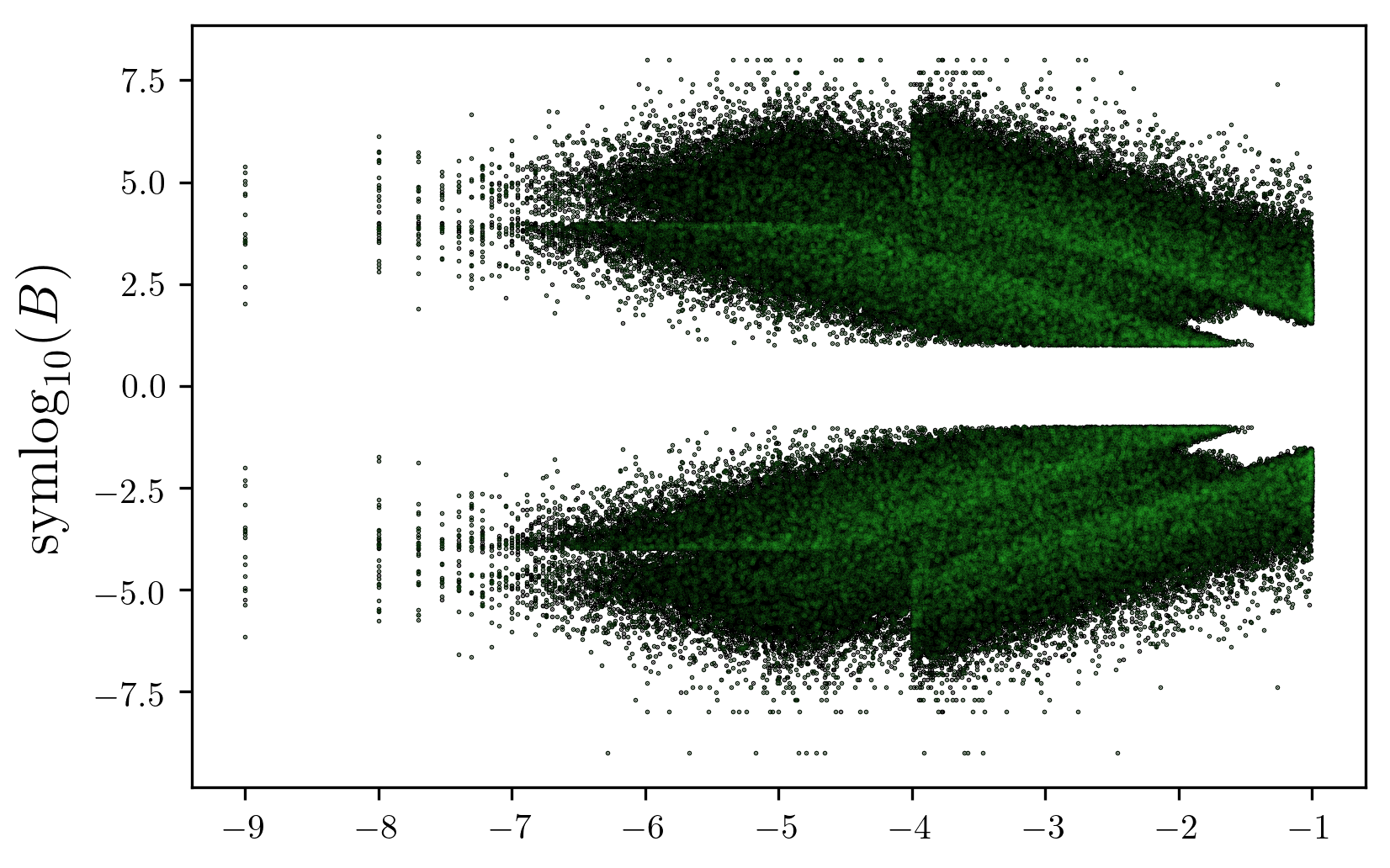}

\includegraphics[width=0.33\linewidth]{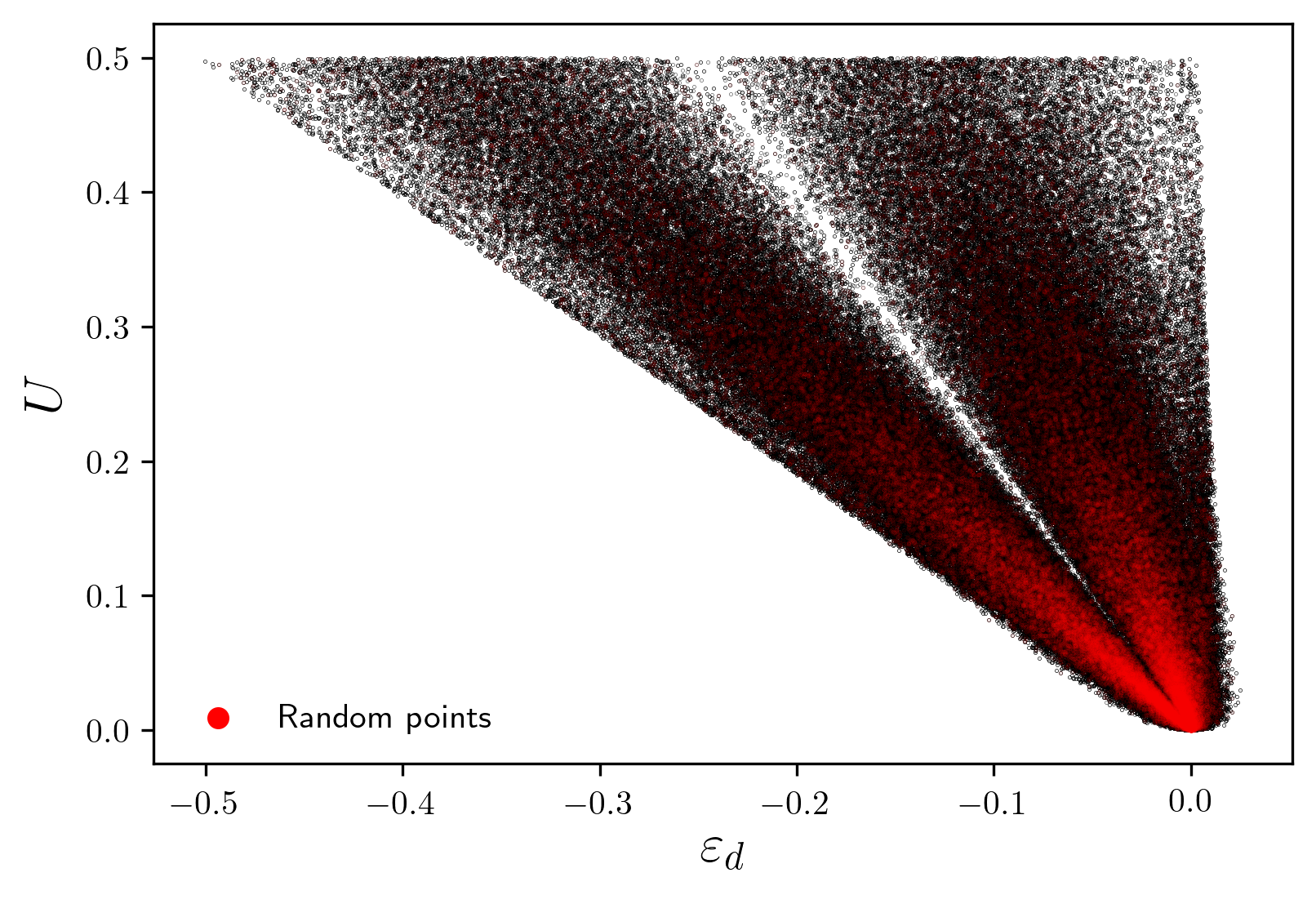}%
\includegraphics[width=0.33\linewidth]{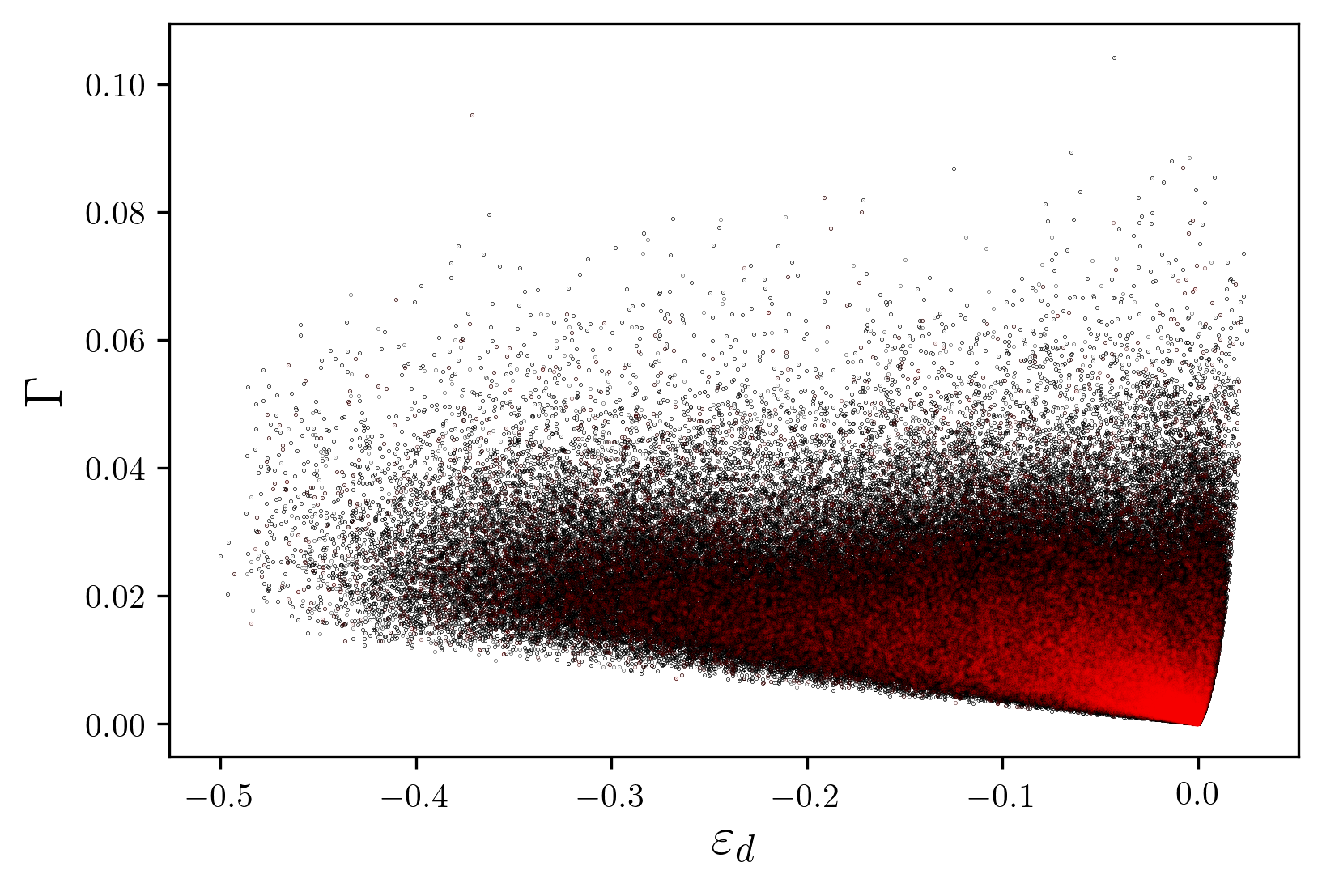}
\includegraphics[width=0.33\linewidth]{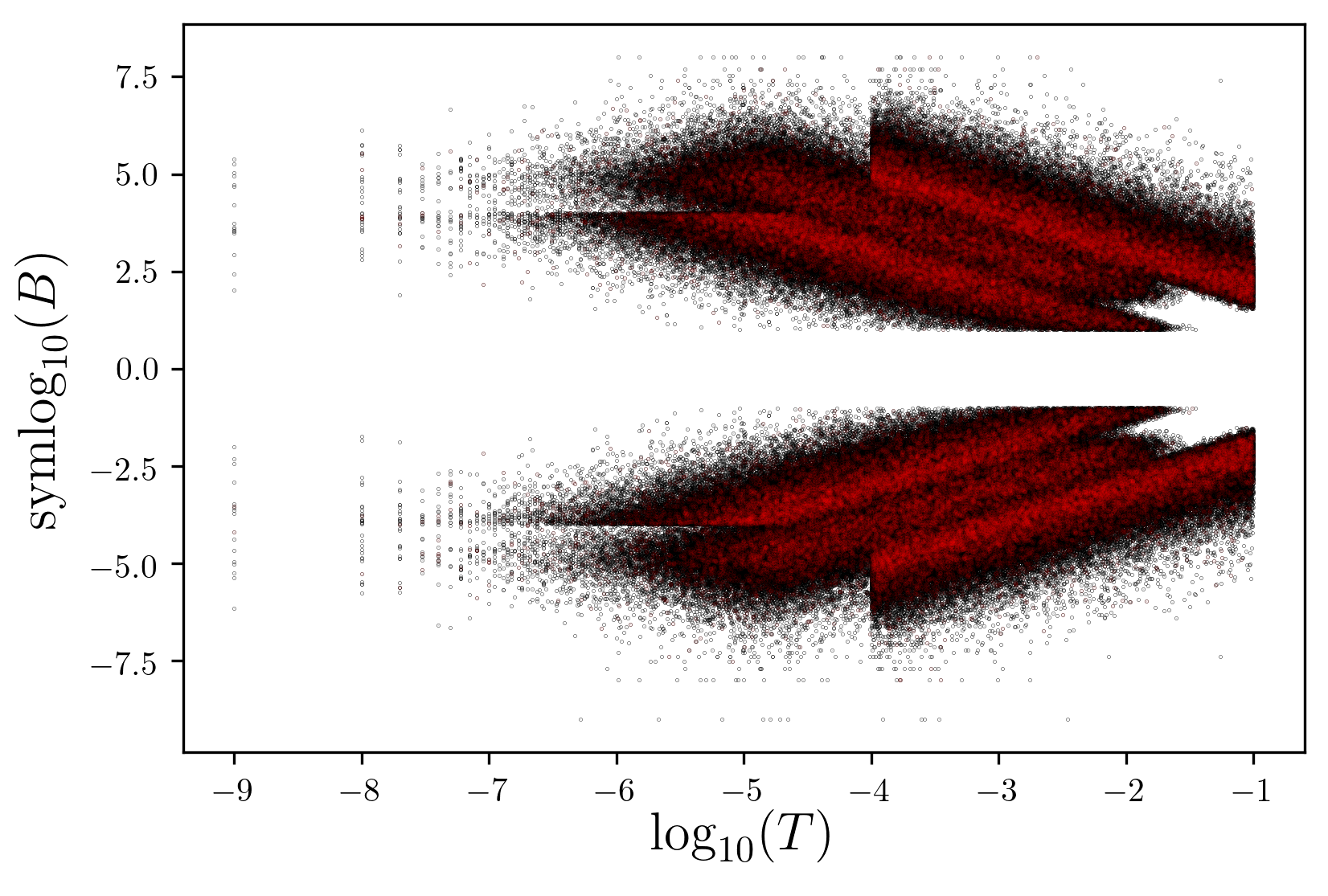}
\caption{
The input feature distribution for the full Anderson set is presented in black. Elements from the down-sampled subsets of the furthest point and random point algorithms each representing approximately 10\% of the full set are superimposed in color at 10\% opacity to exhibit trial density. The top row shows the furthest point distribution in green while the bottom row shows the random points in red. 
See Sections A1.3 and
A1.4 for additional information regarding the
symmetric logarithm (symlog$_{10}$)
procedure and furthest-point sampling algorithm respectively.}
\label{fig:Anderson_parameter_distribution}
\end{figure*}

\begin{figure*}
\centering
\includegraphics[width=0.33\linewidth]{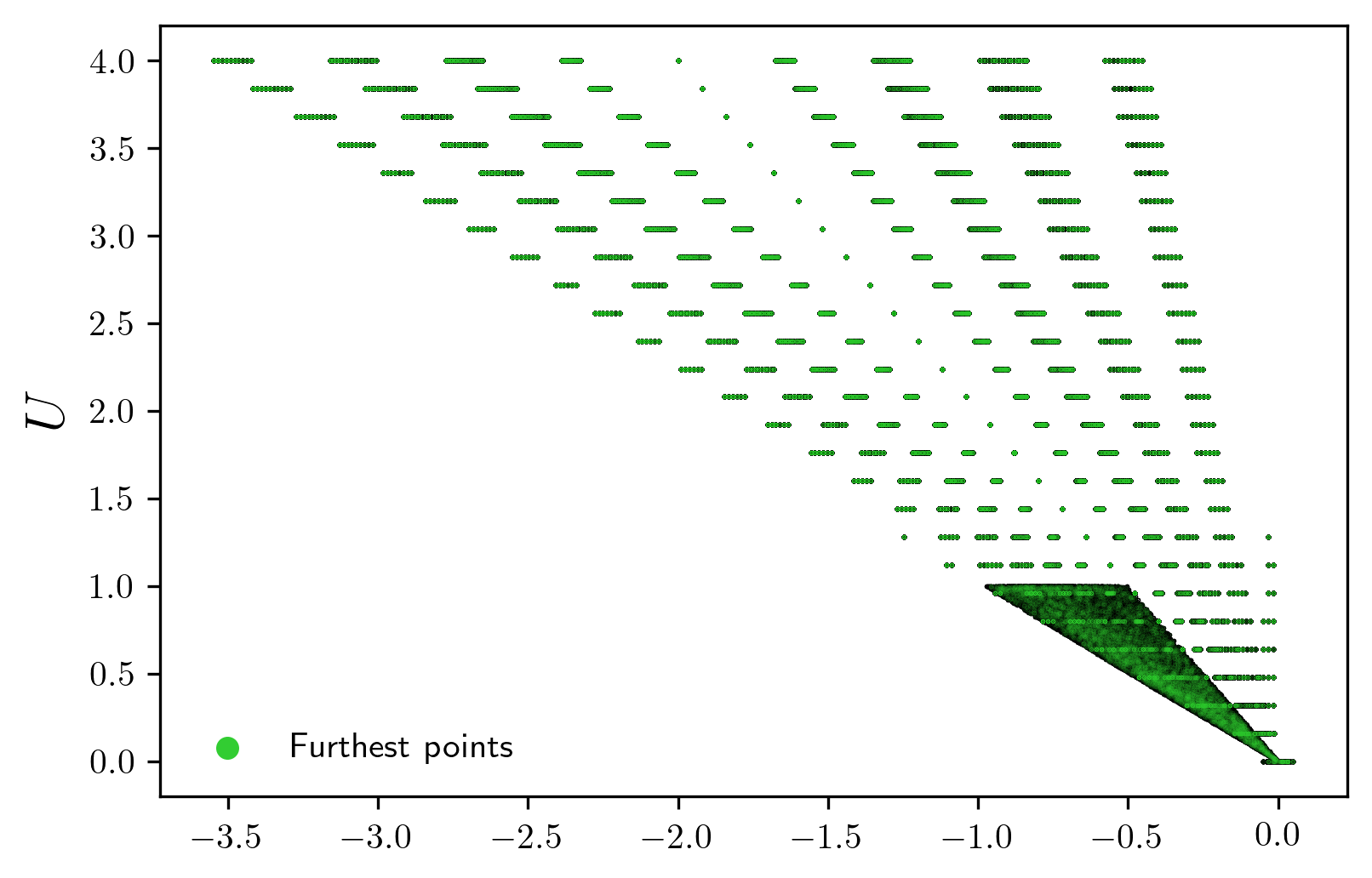}%
\includegraphics[width=0.33\linewidth]{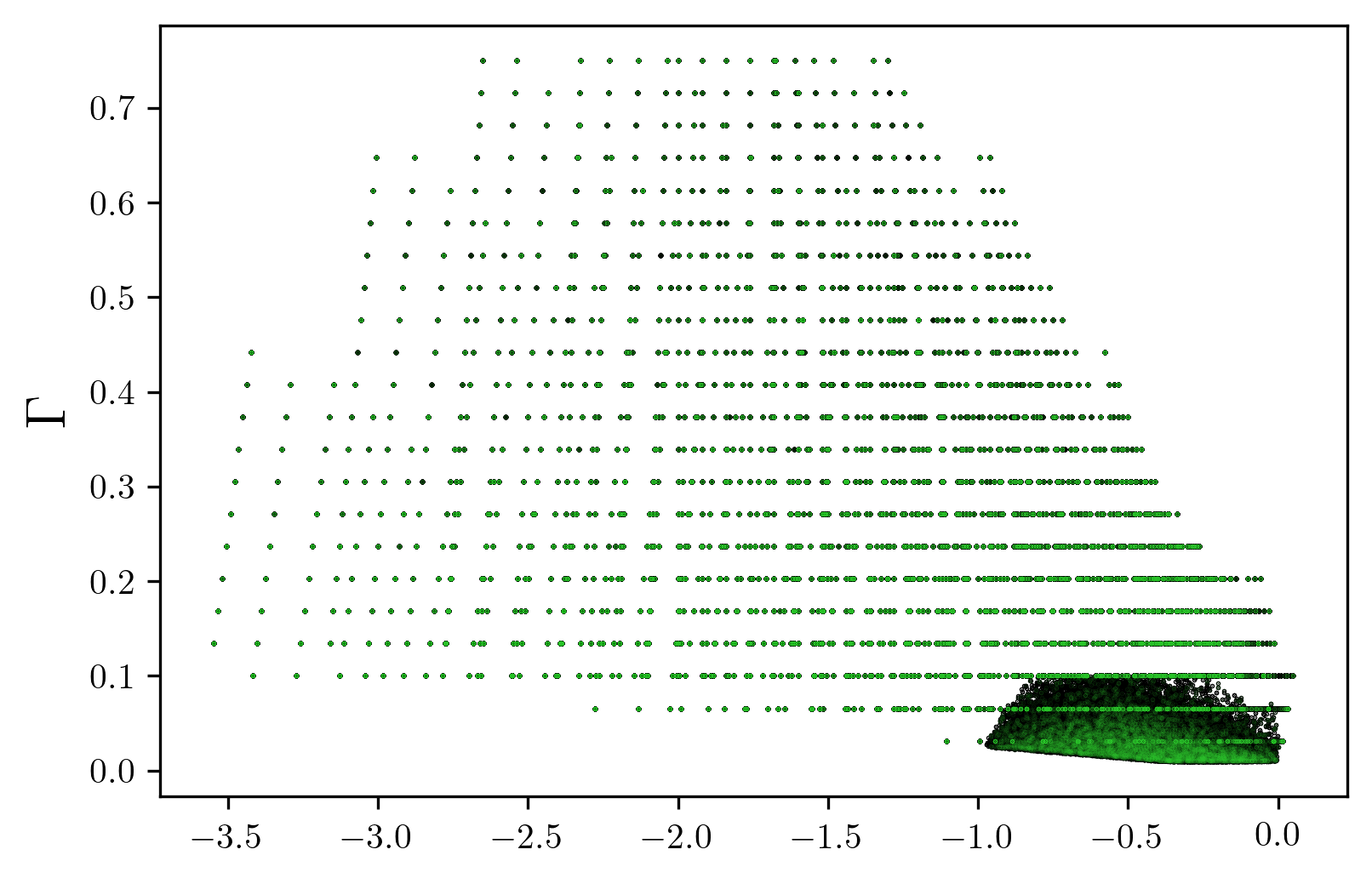}
\includegraphics[width=0.33\linewidth]{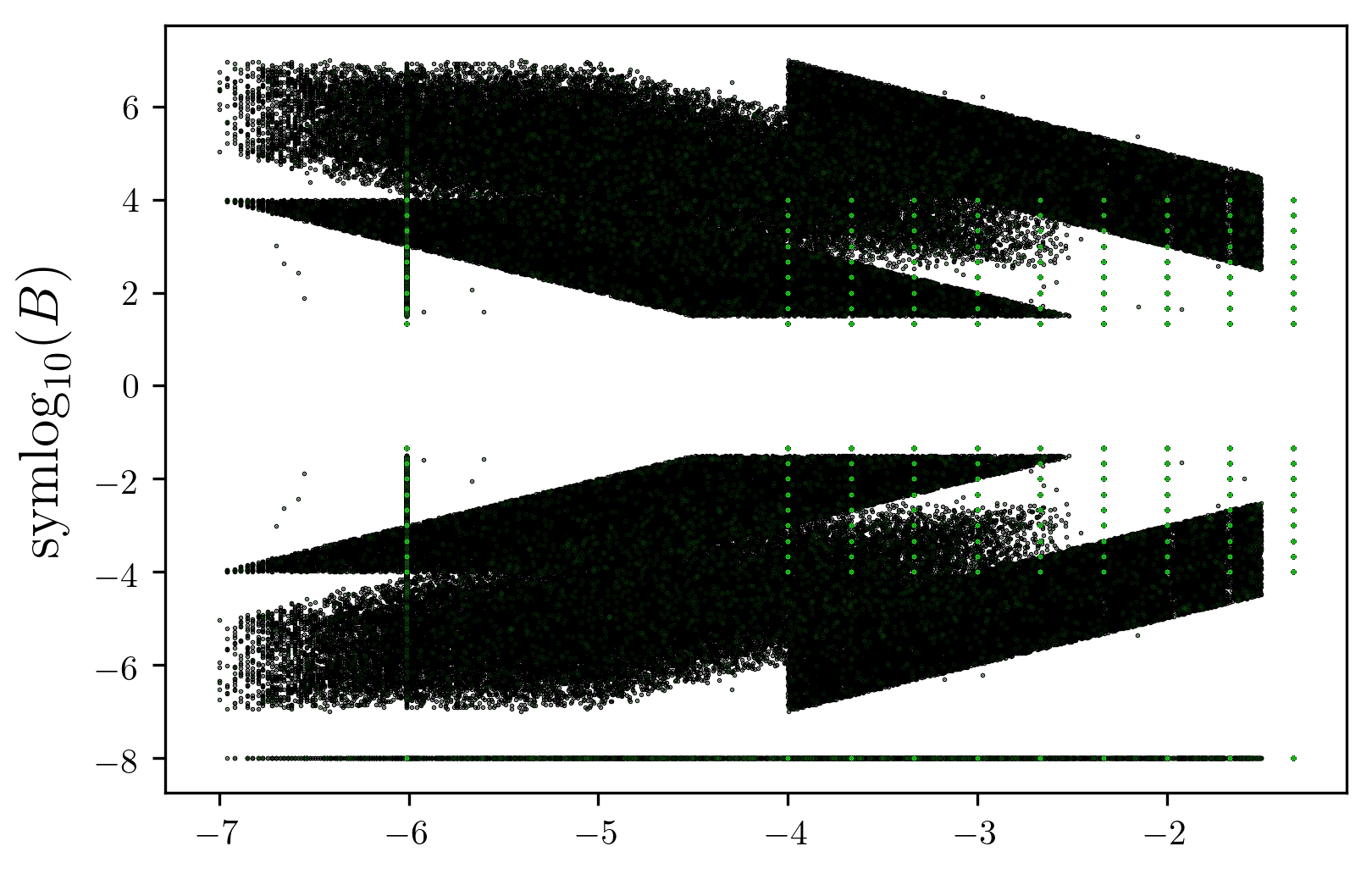}

\includegraphics[width=0.33\linewidth]{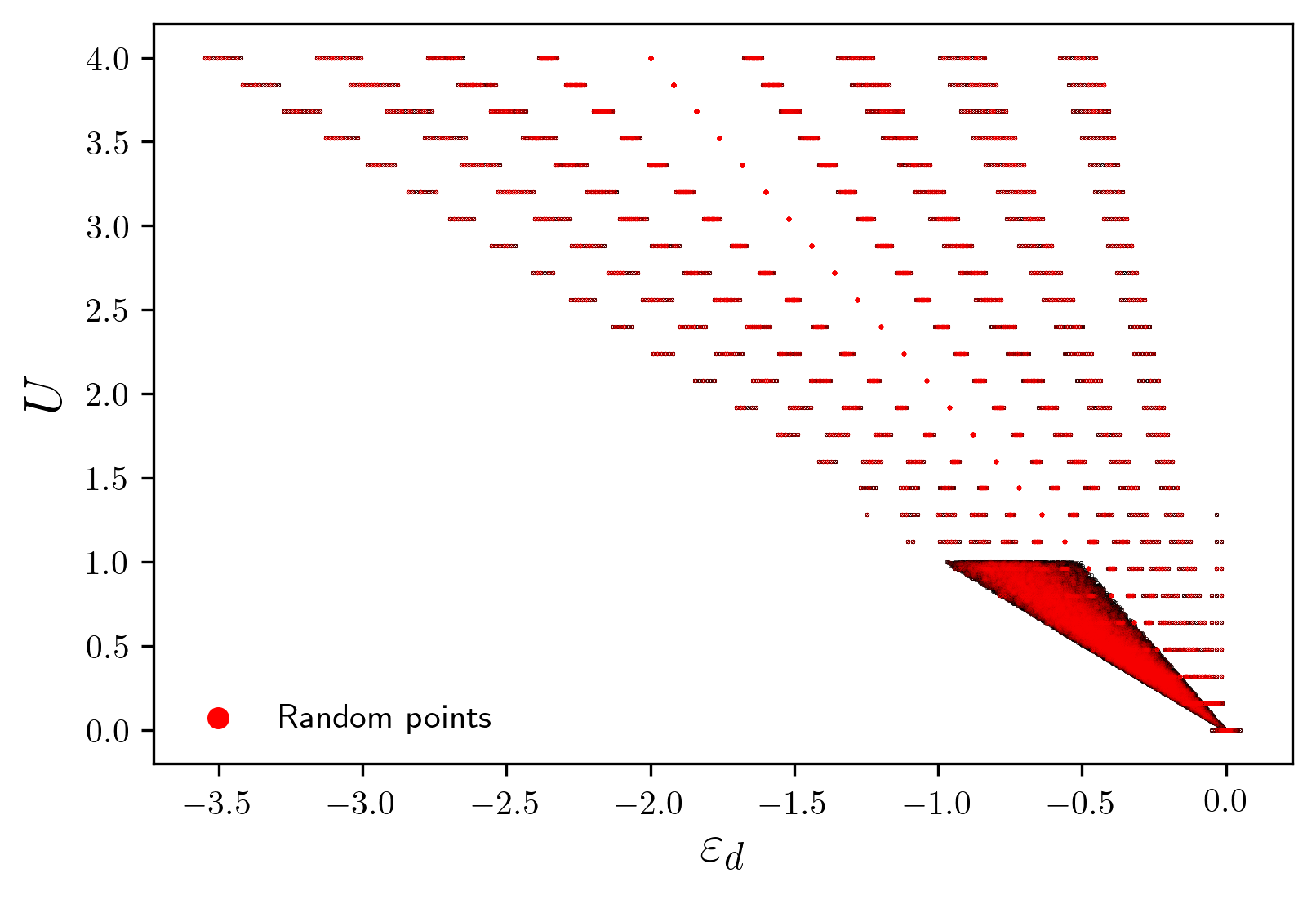}%
\includegraphics[width=0.33\linewidth]{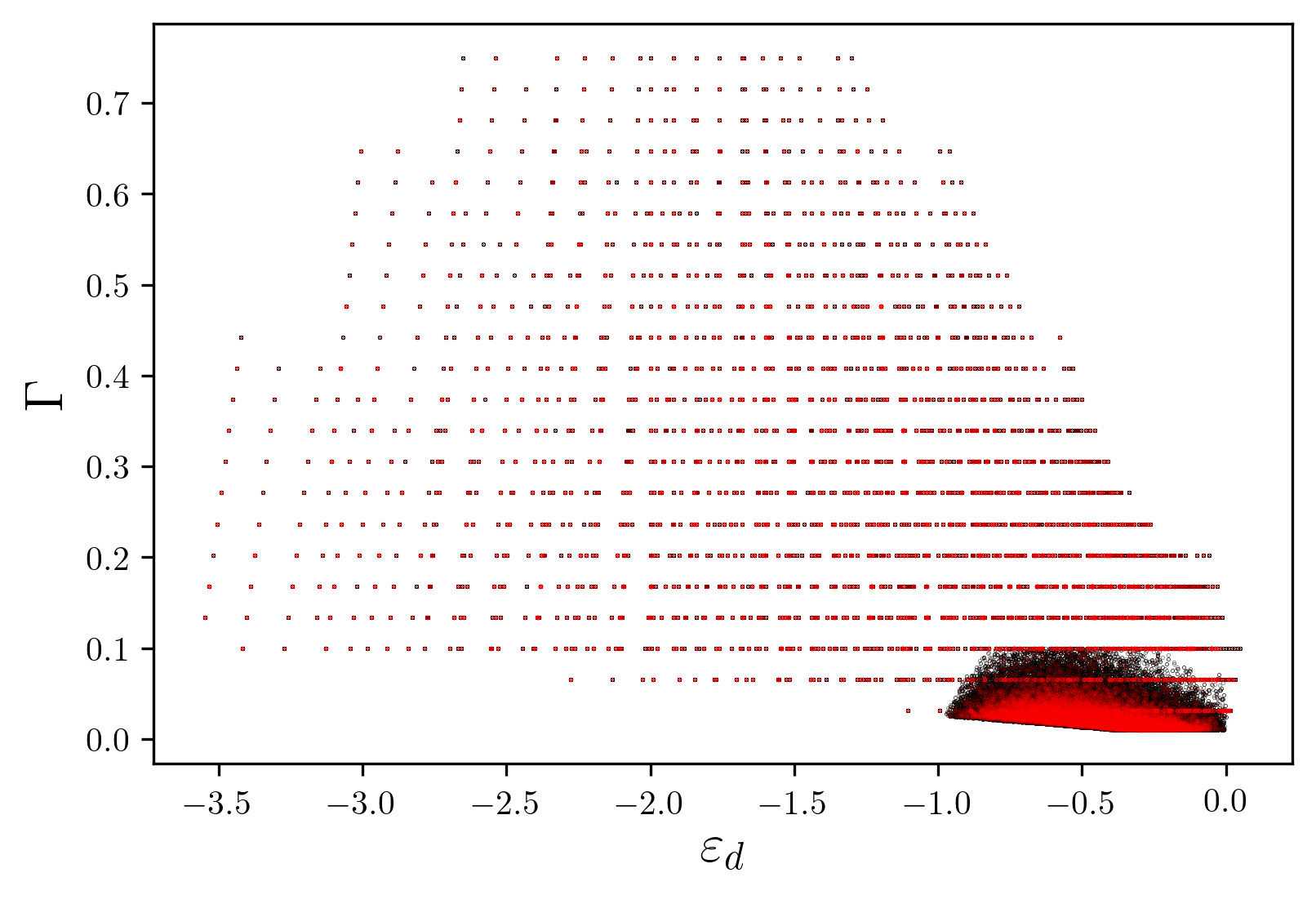}
\includegraphics[width=0.33\linewidth]{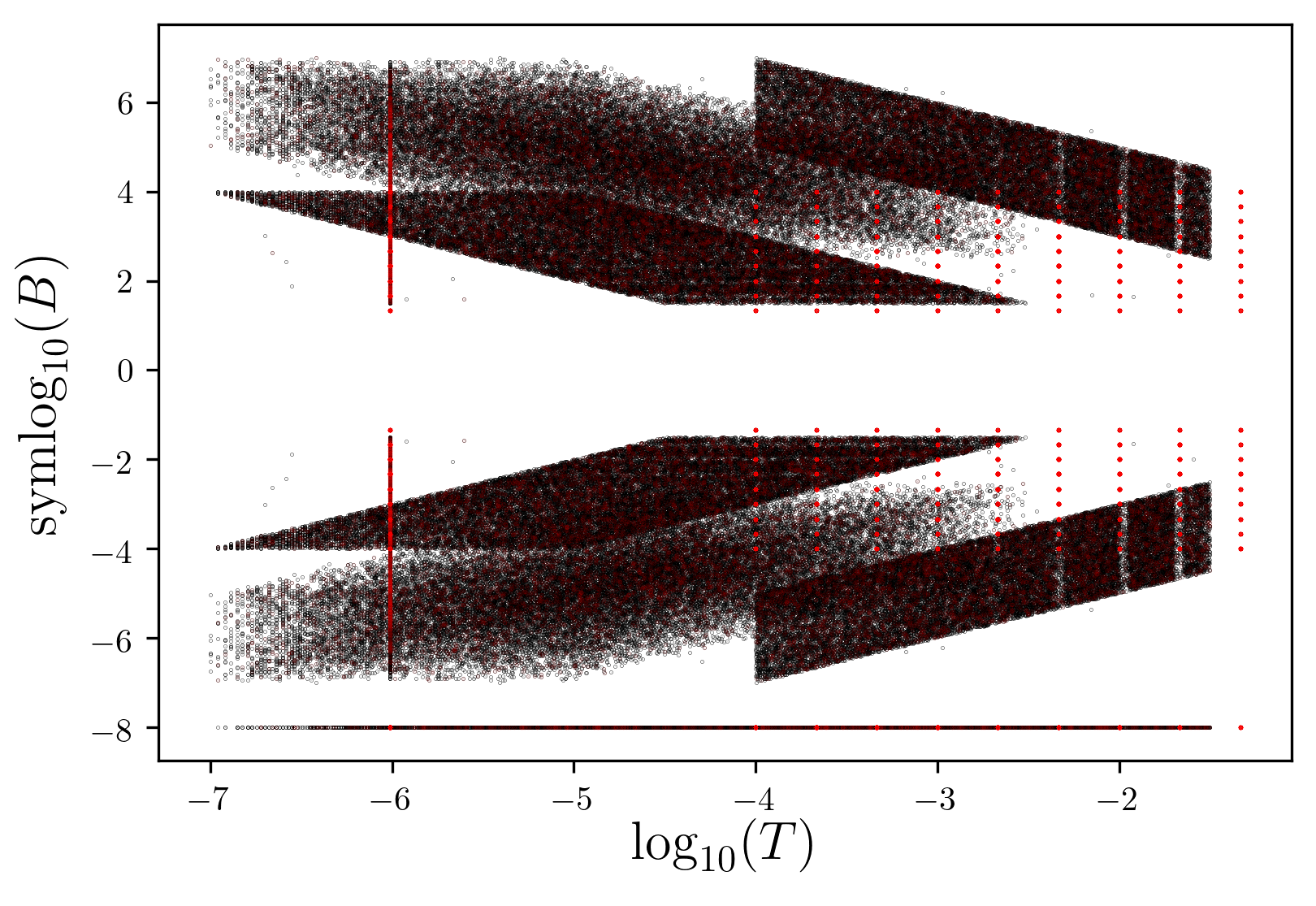}

\caption{ The input feature distribution for the full Kondo
set (black). The down-sampled subsets each representing about
10\% of the full data set  are superimposed in color at 10\%
opacity to demonstrate density. The top row shows the furthest point algorithm's subset in green, while the bottom row has the random point subset in red. See Sections A1.3 and A1.4 for additional information regarding the symmetric logarithm
(symlog$_{10}$) procedure and furthest-point sampling algorithm
respectively. The horizontal
line at symlog$_{10}(B) = -8$ in the right-most panels
is caused by setting our $B\equiv0 \to 10^{-8}$ as explained in Section A1.3.}
\label{fig:Kondo_parameter_distribution}
\end{figure*}

In this section we present the input feature distributions.
In Fig.~\ref{fig:Anderson_parameter_distribution} we show the full Anderson set with its two down-sampled subsets which were plotted with a degree of translucency to exhibit the concentration of the down-sampled trials. One can immediately observe that the furthest point sampled data is more diffuse whereas the random points tend to concentrate in several regions in each distribution. 
In Fig.~\ref{fig:Kondo_parameter_distribution} we present the same data for the full Kondo set. Here, we can clearly see the striations due to the `on-grid' trials as opposed to the clusters of ``off-grid" trials explained in Section A1.

\appendix\section{A2: Machine learning models} 
\label{SI:ML_models}
In this section, we provide a brief overview of some machine learning theory,
along with details of the specific model architectures and hyperparameters
chosen in this work. For a detailed overview of practical machine learning technique in
the domains of physics and chemistry, we refer the interested reader to the
overview by Wang \emph{et al.}~\cite{wang2020machine} and for details on
feed-forward neural networks, to the overview by Cheng and Titterington~\cite{cheng1994neural}.
For details regarding the analytical models, see texts by Mohri {\it et al.}~\cite{Mohri2018} and Hastie {\it et al.}~\cite{Hastie2009}. For the particular applications of these methods explored in this work, we refer our readers to Zhang {\it et al.}~\cite{Zhang2013} and You {\it et al.}~\cite{You2018}. 

\appendix\subsection{A2.1: Neural networks}
\label{SI:neural_networks}
In this work we use a fully-connected feed-forward neural network known as a
multi-layer perceptron (MLP) as the sole deep learning model.
An MLP consists of sequential layers of nodes, where each node takes as input the
output from all previous layers, and outputs a single value, known as an
activation. The activations of the $l$th layer are computed by the general matrix equation
$\veca^l = f(W^l\veca^{l-1}),$
where $\veca^l$ is the vector of activations from the $l$th layer. 
$W^l$ is the $l$th weight matrix mapping the output from the $(l-1)$th layer to the
input of the $l$th layer. 
The function $f$ is called the activation function.
It applies a differentiable, element-wise non-linearity to the output, allowing the 
weights to learn highly-nonlinear representations of the input, parameterized by the
neural network weights. These weights are learned during training, in which a numerical
optimization procedure tries to find the minimum distance between ground truth training
data points and the predictions. The specific model hyperparameters, or non-learned parameters
of the model, are discussed in Section A2.2.

\appendix\subsection{A2.2: Neural network training}
\label{sec:NN_training}
Here we present the training protocols
we used to train and
hyperparameter-tune the neural networks used in this work.
We find the optimal set of hyperparameters
by using a combination of
grid search and hand-tuning. The best model is determined by training models with
different hyperparameter combinations on the same
training set, evaluating on
the validation set $\mathcal{V}$, and
selecting the model with the lowest MAE.\footnote{We remind the reader that the validation and testing sets are fixed for each $\mathcal{D}^\Omega$ for the entirety of this work.}
That model, finally, is evaluated on the testing set
$\mathcal{T}$ which yields the presented data,
unless indicated otherwise.

\appendix\subsection{A2.3: Common neural network training parameters}
\label{SI:NN_training_parameters}
All models share the following hyperparameters:
\begin{itemize}
    \item $||\cdot||_{L_1}$ loss, the mean absolute error (MAE).
    \item Adam optimizer~\cite{kingma2014adam} with a starting learning rate of $10^{-3}.$
    \item ReLU activation function.
    \item A scheduler which decreases the learning rate when the validation loss plateaus. This scheduler has a
    patience of 10 epochs, decrease factor of
    0.5, and minimum learning rate of $10^{-7}.$
    \item Training batch size of 16\,384 ($2^{14}$).
    \item For simplicity, we constrain all hidden layers to be of the
    same size.
\item Total of 5000 epochs.
\end{itemize}

\appendix\subsection{A2.4: Best neural network hyperparameters} 
\label{SI:Best_NN_hyp-params}
As evaluated on $\mathcal{V},$ we present the best hyperparameter
combinations in Table~\ref{SI:tab:bestHP}.
Note that hyperparameter tuning is a highly non-convex problem, and it is
possible that better combinations exist.

\begin{table*}[!htb]
\caption{\label{SI:tab:bestHP} The best hyperparameters for the neural networks for Anderson (upper rows) and Kondo (lower rows) as found by a combination of hand tuning and grid search. Training and validation losses are displayed in Fig.~\ref{fig:anderson losses} and Fig.~\ref{fig:kondo losses} for the Anderson and Kondo sets respectively. Testing set error distributions are shown in Fig.~\ref{fig:anderson hist} and Fig.~\ref{fig:kondo hist} for the Anderson and Kondo sets respectively. In the rightmost column (Examples), we provide the figure number where randomly drawn samples from each of the respective testing sets may be viewed. Both NRG ground truth (black) and predicted (red) spectra are displayed. Summaries of where the example results are shown are given in Tables~\ref{SI:tab:fig res Anderson} and~\ref{SI:tab:fig res Kondo}.}
\begin{ruledtabular}
\begin{tabular}{lccc|c}
Training set &
hidden layer size & Number of hidden layers & Dropout~\cite{srivastava2014dropout} & Examples \\
\colrule
$\mathcal{R}^\mathrm{A}$ 
& 256 & 8 & 0 & \ref{fig:anderson 25 full} \\
$\mathcal{R}_\mathrm{r}^\mathrm{A}$ 
& 256 & 4 & 0.05 & \ref{fig:anderson 25 full R} \\
$\mathcal{R}_\mathrm{f}^\mathrm{A}$ 
& 256 & 8 & 0 & \ref{fig:anderson 25 full F} \\
\colrule
$\mathcal{R}^\mathrm{K}$ 
& 256 & 8 & 0 & \ref{fig:kondo 25 full} \\
$\mathcal{R}_\mathrm{r}^\mathrm{K}$ 
& 256 & 4 & 0.05 & \ref{fig:kondo 25 full R} \\
$\mathcal{R}_\mathrm{f}^\mathrm{K}$ 
& 256 & 8 & 0 & \ref{fig:kondo 25 full F}
\end{tabular}
\end{ruledtabular}
\end{table*}

\appendix\subsection{A2.5: Kernel ridge regression} 
\label{SI:KRR}
Regression methods are trained to find the line of best fit.
However, not all problems lend themselves well to a
linear-fit decision boundary. In those cases it may be
easier to compute the parameters' higher dimensional dual
space by way of a kernel where a linear fit may now be
possible~\cite{Mohri2018}. This approach is known as kernel ridge regression (KRR)~\cite{You2018, scikit-learn}.

For a given trial $\left(x_p, y_p\right)$ of $N$ total trials, the generic form of a KRR minimization algorithm reads: 

\begin{subequations} \label{eqn:generic_KRR}
\begin{align}
  \min\limits_{\alpha} \frac{1}{N}&\left( 
  \sum_{p=1}^{N} \left(y_p - f_p\right)^2
+ \lambda \lVert f \rVert_\mathcal{H}^2
  \right)
\label{eqn:generic_KRR:1}
\\
  f_p &= \sum_{p'=1}^N \alpha_{p'}k\left(x_{p'}, x_p\right)
\label{eqn:generic_KRR:2}
\end{align}
\end{subequations}
where the first term in Eq.~\ref{eqn:generic_KRR:1} is the usual linear regression mean
squared error (MSE) cost function between the model's
kernel-based prediction $f_p$ and ground truth value $y_p$~\cite{Zhang2013}.
While several kernels
exist, in this work we exclusively use the
Laplacian kernel \cite{Rupp15}
given by 
$ K_{pp'} \equiv k( x_p, x_{p'} )
  = \mathrm{exp}\bigl(
       -\tfrac{1}{\sigma}
        \Vert x_p-x_{p'}\Vert_1 
    \bigr)$ 
with input feature vectors $x_p$ and $x_{p'}$
(5-dimensional in this work).
The exponential argument is the L1 norm
(Manhattan Distance) divided by the kernel radius
$\sigma > 0$
which determines how similar $x_p$ is to $x_{p'}$. 
The target vectors $y_p$ are of size $M=333$ 
in the present case.

The second term in \Eq{eqn:generic_KRR:1}
is the regularization term which helps prevent over-fitting.
It includes two factors: the strength
$\lambda \geq 0$, and the Hilbert space norm $\lVert f
\rVert_\mathcal{H}$ defined as $\lVert f \rVert_\mathcal{H} :=
\langle f,f \rangle_\mathcal{H}^{\sfrac{1}{2}}$.~\cite{Zhang2013} Both the
kernel radius $\sigma$ and regularization strength $\lambda$
are tunable hyperparameters. 
Table~\ref{tab:KRR-hyp-param} shows the selected hand-tuned
hyperparameters.
In this work we find that scaling the data by a
symmetric logarithm procedure for $B$, $T$, and $\Gamma$
only improved the analytical algorithm results when the model is fit to
$\mathcal{R}_\mathrm{r}$. 
Hence, no such scaling is applied when training with $\mathcal{R}_\mathrm{f}$.
Throughout, we apply the standard normalization
of the data, such that
mean and standard deviations of all training data
are 0 and 1, respectively [cf. Section A1.3].

Solving \Eq{eqn:generic_KRR} for the weight matrix \textbf{$\alpha$}
requires an expensive inversion of the kernel matrix, 
$k(x_p, x_{p'})$
which scales as $O(N^3)$ in time and $O(N^2)$ 
in memory for $N$ data points $(x_p,y_p)$.
This expense can be mitigated in several 
ways, including the divide-and-conquer approach, 
detailed below. In this work all KRR trials were modelled on 50k 
training trials (see Section A3).

The $r^2$ score (R-squared) value determines the
quality of a trained regression model. A
perfect model would achieve $r^2=1$,
so the closer a model is to 1, the better the fit.
A score of $r^2=0$
indicates that a constant model predicts the same result
regardless of input. A negative score indicates that the
model is arbitrarily worse than a constant representing the mean
value. In this work we
found that it was possible to generate acceptable models for
the furthest-point sampled training data $\mathcal{R}_\mathrm{f}$, 
but the models fit with random point sampled $\mathcal{R}_\mathrm{r}$
training set were so poor that they had negative $r^2$ values.
Therefore, if training set size is a limiting factor for a brute-force affordable KRR model (here inverting 
a full matrix of dimension 50k),
then one must chose the training set with great care.

\appendix\subsection{A2.6: Divide-and-conquer kernel ridge regression}
\label{SI:DC-KRR}
Divide-and-conquer kernel ridge regression (DC-KRR) is one
of several approaches one may take to mitigate the poor
scaling of a traditional KRR algorithm explained above~\cite{Zhang2013}. In
DC-KRR the total number $N\approx 500k$ of training trials
is subdivided into $S\sim10$ 
subsets with an equal number $n\approx 50k$ 
training trials in each
(with minor variations for the Kondo as compared to
the Anderson set). Then, a 
separate KRR model is fitted for each, and the trained weights
($\alpha$ in \Eq{eqn:generic_KRR}) are saved. After training, a 
prediction may be acquired by computing (for each subset) the product of the weights and the kernel of the queried value 
and subset's training trials. The final results are averaged for the final prediction as shown in Eq.~\ref{Eqn:DC-KRR_prediction}.

Given input parameters $x_P$ corresponding to ground truth target $y_P$, one may obtain a DC-KRR prediction $\hat{y}_P$ via:
\begin{equation}
    \label{Eqn:DC-KRR_prediction}
    \hat{y}_P = \frac{1}{S}\sum_{s=1}^S \mathrm{ker}(\mathcal{R}_f^s, x_P) \alpha_s
\end{equation}
where $\alpha_s$ is a matrix of the learned algorithm weights described in Eqn.~\ref{eqn:generic_KRR} and ker$(\mathcal{R}_f^s, x_P)$ indicates the kernel between all $x_p \in \mathcal{R}_f^s$ and the queried $x_P$ input. 
DC-KRR has hyperparameters $\lambda$ and $\sigma$ which function in the same manner as described in Section 2.5 and we use the Laplacian kernel here as well. Final hyperparameter values are given in Table~\ref{tab:KRR-hyp-param}, which are applied universally for all subsets. Fig.~\ref{fig:DC-KRR:r2} shows the individual $r^2$ values for each subset in the DC-KRR algorithm, whereas the final averaged $r^2$ value is reported in Table~\ref{tab:KRR-hyp-param}.

This approach reduces the cost of training with 
all data points in $\mathcal{R}$ and making the approach easily 
parallelizable. However, the overall scaling has not changed in this
implementation; DC-KRR simply permits the use of the full 
training set. Additionally, it should be noted that the composition of 
the individual subsets matters. As DC-KRR works by training a 
multitude of standard KRR models, the same pitfalls that can 
negatively affect a single KRR model can also detract from the 
individual DC-KRR subset models which are then averaged, thus potentially
compounding the error. In this work we chose to apply the furthest-
point ordering on the training data prior to splitting $\mathcal{R}$
set into equal sized subsets as the KRR trials performed better 
with this pre-processing step. 

\appendix\subsection{A2.7: Best KRR and DC-KRR hyperparameters}
\label{SI:Best_KRR_hyp-params}
\begin{table*}
\caption{
   The hand-tuned hyperparameters $\sigma$ and $\lambda$
   [cf. \Eqs{eqn:generic_KRR}], together with the
   $r^2$ scores of the
   validation ($\mathcal{V}$) and testing sets ($\mathcal{T}$) 
   used in this work for kernel ridge
   regression (KRR) and divide-and-conquer KRR (DC-KRR).
   While several values
   were examined for each of these hyperparameters for each set of
   training data, only the final choices are presented.
   We also differentiate furthest-point sampled $\mathcal{R}_\mathrm{f}$
   from random point sampled $\mathcal{R}_\mathrm{r}$ training data.
   For the training of the DC-KRR models, the full training set
   $\mathcal{R}$ was used, yet
   partitioned into $S$ disjoint subsets $\{ \mathcal{R}_f^s \}$ as described in Eqn.~3 in the main text.
   Values for the DC-KRR $r^2$ entries are the result of averaging the $r^2$ values from the subsets shown in Fig.~\ref{fig:DC-KRR:r2}. In the rightmost column (Examples), we provide the figure number where randomly drawn samples from each of the respective testing sets may be viewed. Both NRG ground truth (black) and predicted (red) spectra are displayed. Summaries of where the example results are shown are given in Tables~\ref{SI:tab:fig res Anderson} and~\ref{SI:tab:fig res Kondo}. Only the best models have associated result examples in Appendix A3. Figs.~\ref{fig:anderson hist krr} and~\ref{fig:kondo hist krr} demonstrate regression approach error distributions for the Anderson and Kondo sets respectively.
\label{tab:KRR-hyp-param}
}

\begin{ruledtabular}
\begin{tabular}{cc|cccccc|cccccc}
& & \multicolumn{6}{c|}{\textrm{Anderson}} & \multicolumn{6}{c}{\textrm{Kondo}} \\
\hline
\textbf{KRR} & Scaling &
 $\sigma$ & $\lambda$ & $r^2_{\mathcal{V}}$ & $r^2_{\mathcal{T}}$ & MAE($\mathcal{T}$) & Examples &
 $\sigma$ & $\lambda$ & $r^2_{\mathcal{V}}$ & $r^2_{\mathcal{T}}$ & MAE($\mathcal{T}$) & Examples\\
\hline

$\mathcal{R}_\mathrm{r}$ 
 & None & 100 & 0 & -0.267 & -0.255 & $0.102\pm0.096 $ & N/A & 100 & 0 & -0.192 & -0.204 & $0.182 \pm 0.187$ & N/A \\
 
$\mathcal{R}_\mathrm{r}$ 
 & Symlog & $6.\bar{6}$ & 0.1 & 0.277 & 0.282 & $0.085 \pm 0.068$ & \ref{fig:anderson 25 full R krr}  & 10 & 0 & 0.179 & 0.188 & $0.152 \pm 0.145$ & \ref{fig:kondo 25 full R krr} \\
 
$\mathcal{R}_\mathrm{f}$ 
 & None & 1 & 0.01 & 0.929 & 0.932 & $0.021 \pm 0.023$ & \ref{fig:anderson 25 full F krr} & 1 & 0.01 & 0.979 & 0.979 & $0.019 \pm 0.021$ &  \ref{fig:kondo 25 full F krr}\\
 
$\mathcal{R}_\mathrm{f}$ 
 & Symlog & 1 & 0 & 0.909 & 0.915 & $0.024 \pm 0.026$ & N/A  & 1 & 0.01 & 0.972 & 0.972 & $0.023 \pm0.024$  & N/A \\
\hline
\textbf{DC-KRR}
  & & & & & & \\ 
\hline
$\mathcal{R}_\mathrm{F}$
& None & 1 & 0 & 0.950 & 0.950 & $0.0175 \pm 0.026$& \ref{fig:anderson 25 full krr} & 1 & 0 & 0.902 & 0.899 & $0.035 \pm 0.033$ & \ref{fig:kondo 25 full krr} \\
\end{tabular}
\end{ruledtabular}
\end{table*}

\begin{figure*}
\centering
\includegraphics[width=\columnwidth]{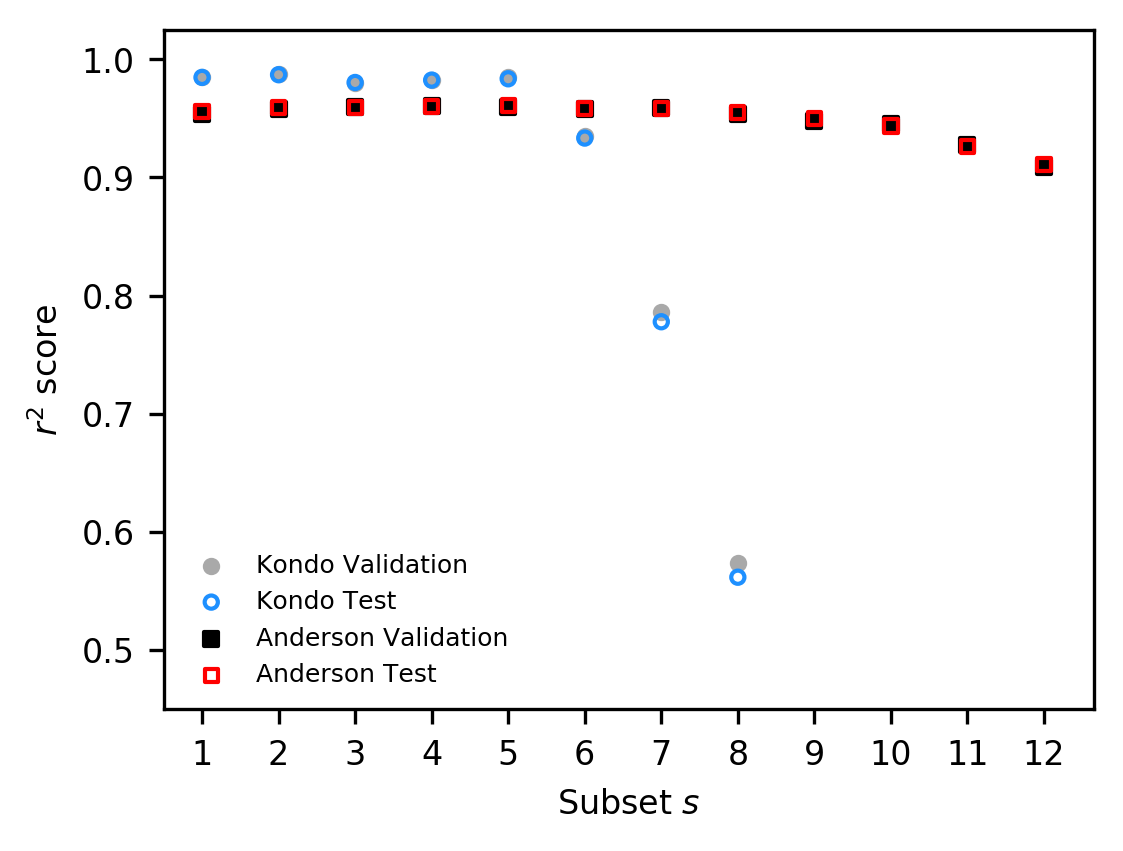}
\caption{DC-KRR $r^2$ values for each subset within the Anderson (squares) and Kondo (circles) $\mathcal{R}_\mathrm{F}$ sets. Closed markers indicate the validation $r^2$ scores, and open markers indicate that of the test sets. Final DC-KRR values are the averages across all relevant subsets.
}
\label{fig:DC-KRR:r2}
\end{figure*}

In \Fig{fig:DC-KRR:r2}
we present the $r^2$ values for each subset $s$ for the validation $\mathcal{V}$ and testing  $\mathcal{T}$ sets in DC-KRR.
In all cases there is excellent agreement between
a given subset's $r^2$ score for the validation and test sets.
By the very construction of the underlying FPS, the 
last group ($s=S$) collects the rest of the data set,
hence is no longer strictly FPS. This introduces a
certain bias towards late patches,
as seen in \Fig{fig:DC-KRR:r2}.
The values for the Anderson set
are tightly distributed. In contrast,
the Kondo sets exhibit significant
deterioration towards 
the last subsets $s=7$ and 8,
negatively impacting the overall 
average. Despite this deterioration, however,
the final $r^2$ values remain quite high for both datasets, indicating reliable models.

\appendix\section{A3: Results}\label{SI:Results}
In this section, we present figures demonstrating various representations of
each model's performance. These figures are connected to the different models in
Table~\ref{SI:tab:fig res Anderson} for the Anderson results, and
Table~\ref{SI:tab:fig res Kondo} for the Kondo results.
For clarity, we highlight the differences between each dataset; note that all results that follow
were hyperparameter-tuned on the same $\mathcal{V}$ and all results that follow in this section
correspond to the testing set ($\mathcal{T}$) results as evaluated on the best model, which is
determined by the MAE on $\mathcal{V}$.
\begin{itemize}

\item The full training set
    refers to training on all available training data,
    $\mathcal{R}$.

\item The FPS training set
    refers to training on only the 50k furthest-points
    sampled data, $\mathcal{R}_\mathrm{f} = \mathcal{R}_\mathrm{f}^{(1)}.$

\item The dataset $\mathcal{R}_\mathrm{r}$
     refers to training on only the 50k randomly
     down-sampled data, 
\end{itemize}

For all datasets, we present the following results:

\begin{itemize}
\item Random 25: a randomly-selected 25 samples plotted on the ML grid.
\item Training info: loss and learning rate plots when applicable.
\item Distribution: error distributions.
\end{itemize}

\begin{table*}[!htb]
\caption{\label{SI:tab:fig res Anderson} A quick-reference
for the \emph{Anderson} results presented in this section.
The table entries are figure labels.
See text for description for the three rows.
}
\begin{ruledtabular}
\begin{tabular}{lcccccc}
&
MLP$(\mathcal{R})$&
MLP$(\mathcal{R}_\mathrm{r})$&
MLP$(\mathcal{R}_\mathrm{f})$&
KRR$(\mathcal{R}_\mathrm{r})$&
KRR$(\mathcal{R}_\mathrm{f})$&
DC-KRR$(\mathcal{R}_\mathrm{F})$\\
\colrule
Random 25 &  \ref{fig:anderson 25 full} & \ref{fig:anderson 25 full R} & \ref{fig:anderson 25 full F} & \ref{fig:anderson 25 full R krr} & \ref{fig:anderson 25 full F krr} & \ref{fig:anderson 25 full krr} \\
Training info & \ref{fig:anderson losses} & \ref{fig:anderson losses} & \ref{fig:anderson losses} & $-$ & $-$ & $-$ \\
Distribution & \ref{fig:anderson hist} & \ref{fig:anderson hist} & \ref{fig:anderson hist}
& \ref{fig:anderson hist krr} & \ref{fig:anderson hist krr} & \ref{fig:anderson hist krr}
\end{tabular}
\end{ruledtabular}
\end{table*}

\begin{table*}[!htb]
\caption{\label{SI:tab:fig res Kondo} A quick-reference for the \emph{Kondo} results presented in this section.
The table entries are figure labels.
See text for description fo the three rows.
}
\begin{ruledtabular}
\begin{tabular}{lcccccc}
&
MLP$(\mathcal{R})$&
MLP$(\mathcal{R}_\mathrm{r})$&
MLP$(\mathcal{R}_\mathrm{f})$&
KRR$(\mathcal{R}_\mathrm{r})$&
KRR$(\mathcal{R}_\mathrm{f})$&
DC-KRR$(\mathcal{R}_\mathrm{F})$\\
\colrule
Random 25 & \ref{fig:kondo 25 full} & \ref{fig:kondo 25 full R} & \ref{fig:kondo 25 full F} & \ref{fig:kondo 25 full R krr} & \ref{fig:kondo 25 full F krr}
& \ref{fig:kondo 25 full krr} \\
Training info & \ref{fig:kondo losses} & \ref{fig:kondo losses} & \ref{fig:kondo losses} & $-$ & $-$ & $-$ \\
Distribution & \ref{fig:kondo hist} & \ref{fig:kondo hist} & \ref{fig:kondo hist}
& \ref{fig:kondo hist krr} & \ref{fig:kondo hist krr} & \ref{fig:kondo hist krr}
\end{tabular}
\end{ruledtabular}
\end{table*}

\begin{figure*}[p] 
\centering
\includegraphics[width=\linewidth]{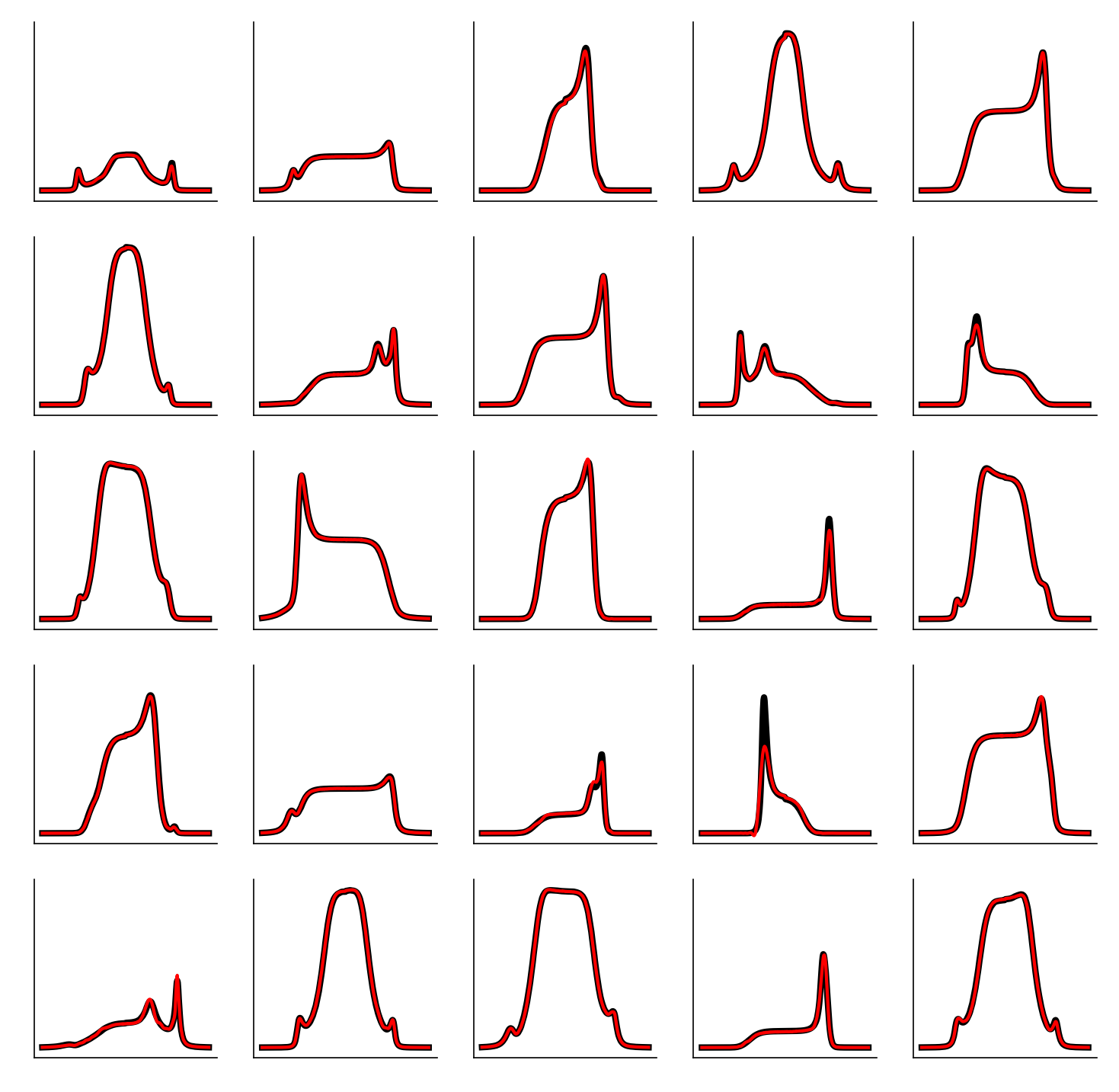}
\caption{\label{fig:anderson 25 full} Random samples on the Anderson testing set for an MLP trained on $\mathcal{R}.$ Black is ground truth and red is prediction.}
\end{figure*}

\begin{figure*}[p] 
\centering
\includegraphics[width=\linewidth]{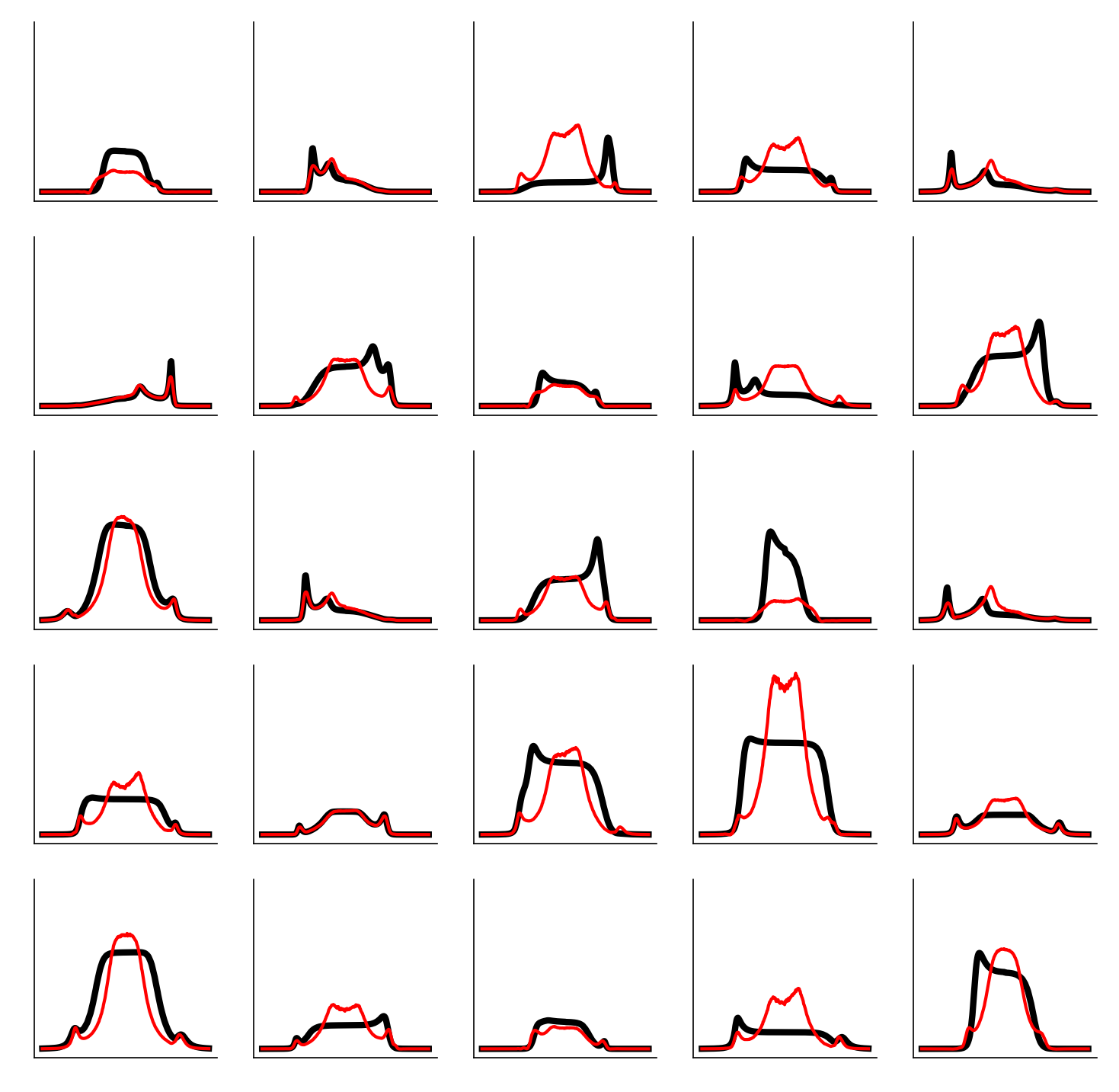}
\caption{\label{fig:anderson 25 full R} Random samples on the Anderson testing set for an MLP trained on $\mathcal{R}_\mathrm{r}.$ Black is ground truth and red is prediction.}
\end{figure*}

\begin{figure*}[p] 
\centering
\includegraphics[width=\linewidth]{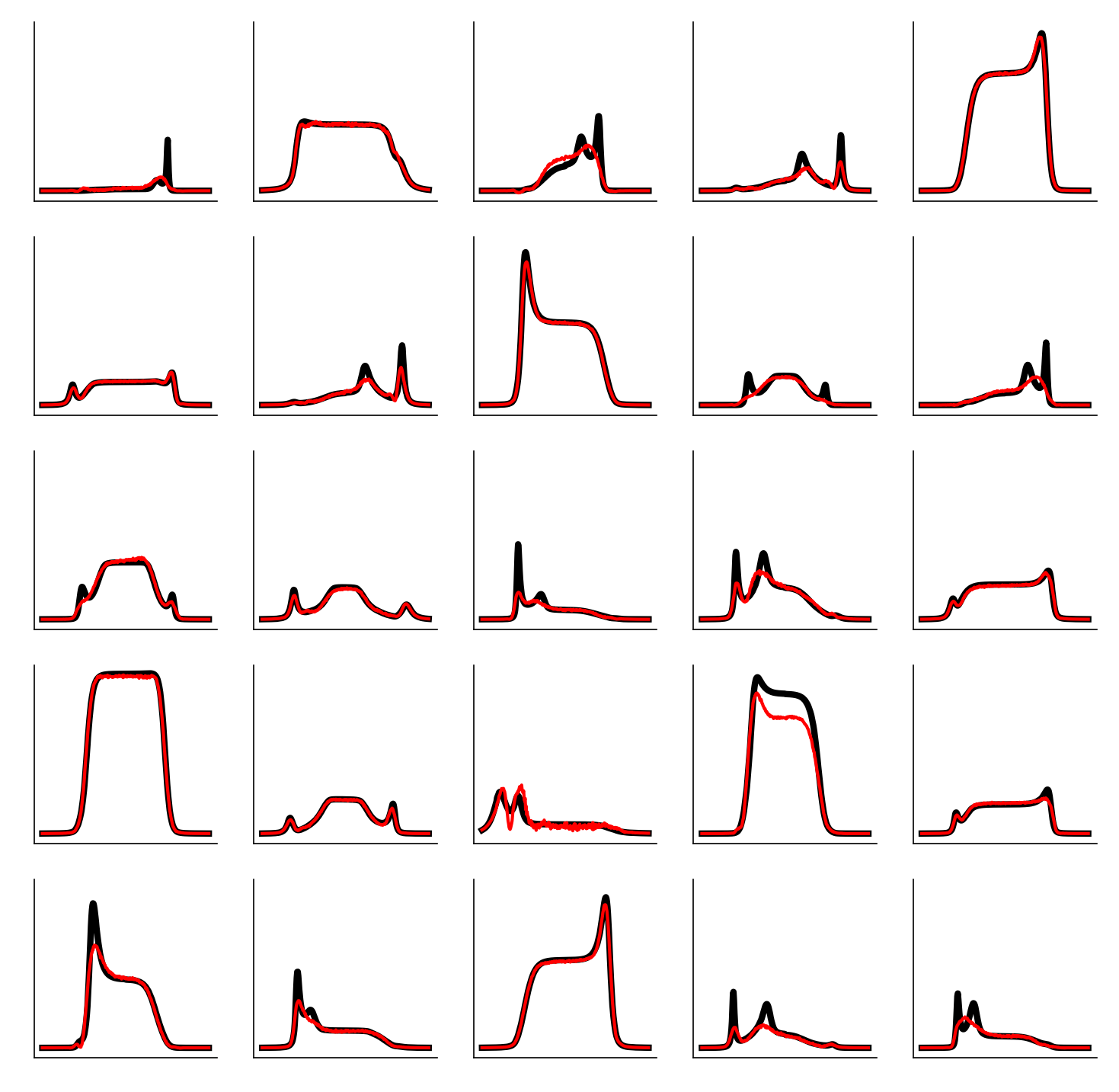}
\caption{\label{fig:anderson 25 full F} Random samples on the Anderson testing set for an MLP trained on $\mathcal{R}_\mathrm{f}.$ Black is ground truth and red is prediction.}
\end{figure*}

\begin{figure*}[p] 
\centering
\includegraphics[width=\linewidth]{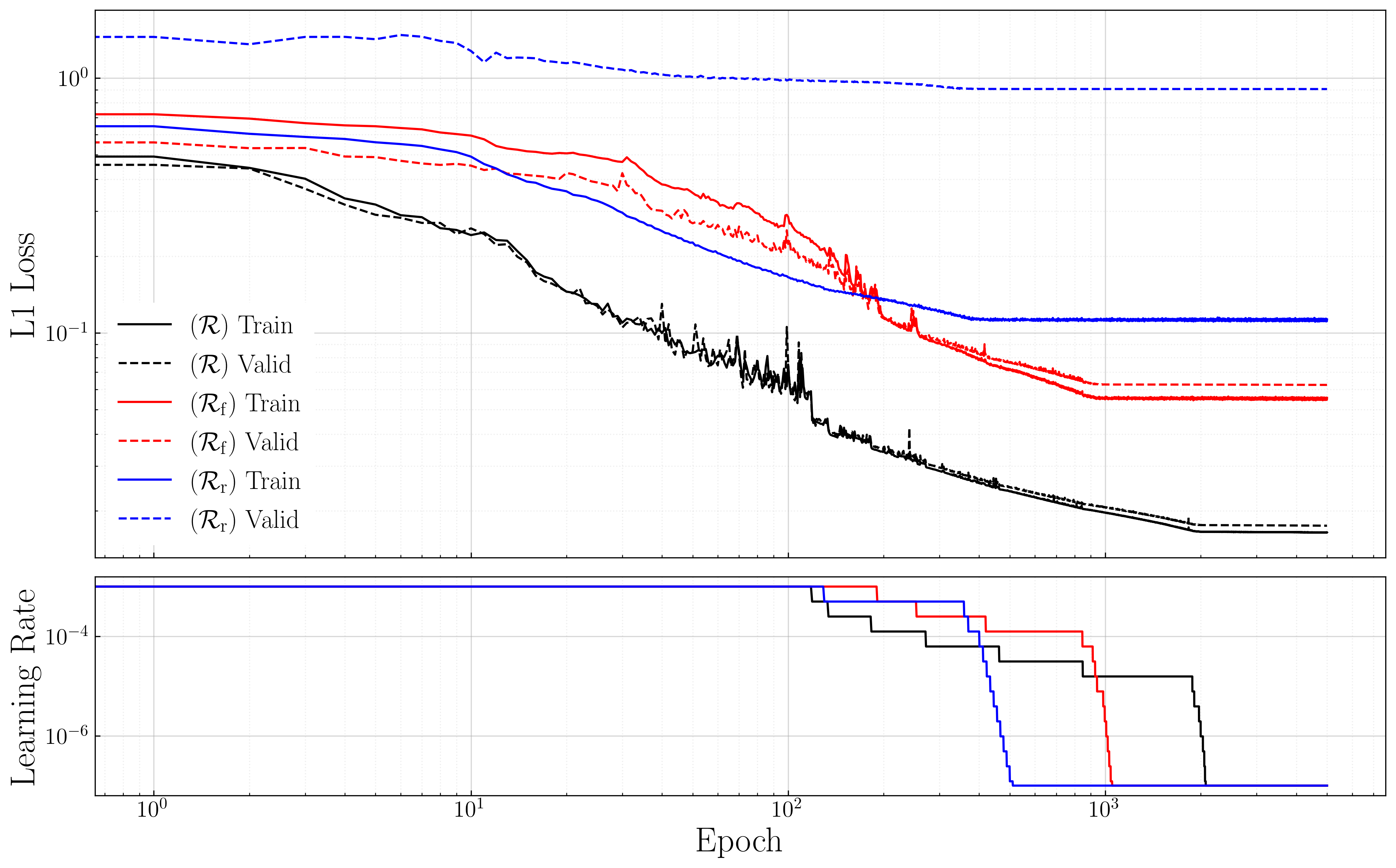}
\caption{\label{fig:anderson losses} Anderson training and validation losses and learning rates
plotted as a function of epochs for models trained on the full set $\mathcal{R},$ the FPS training set
$\mathcal{R}_\mathrm{f},$ and random-sampled $\mathcal{R}_\mathrm{r},$ training sets.}
\end{figure*}

\begin{figure*}[p] 
\centering
\includegraphics[width=\linewidth]{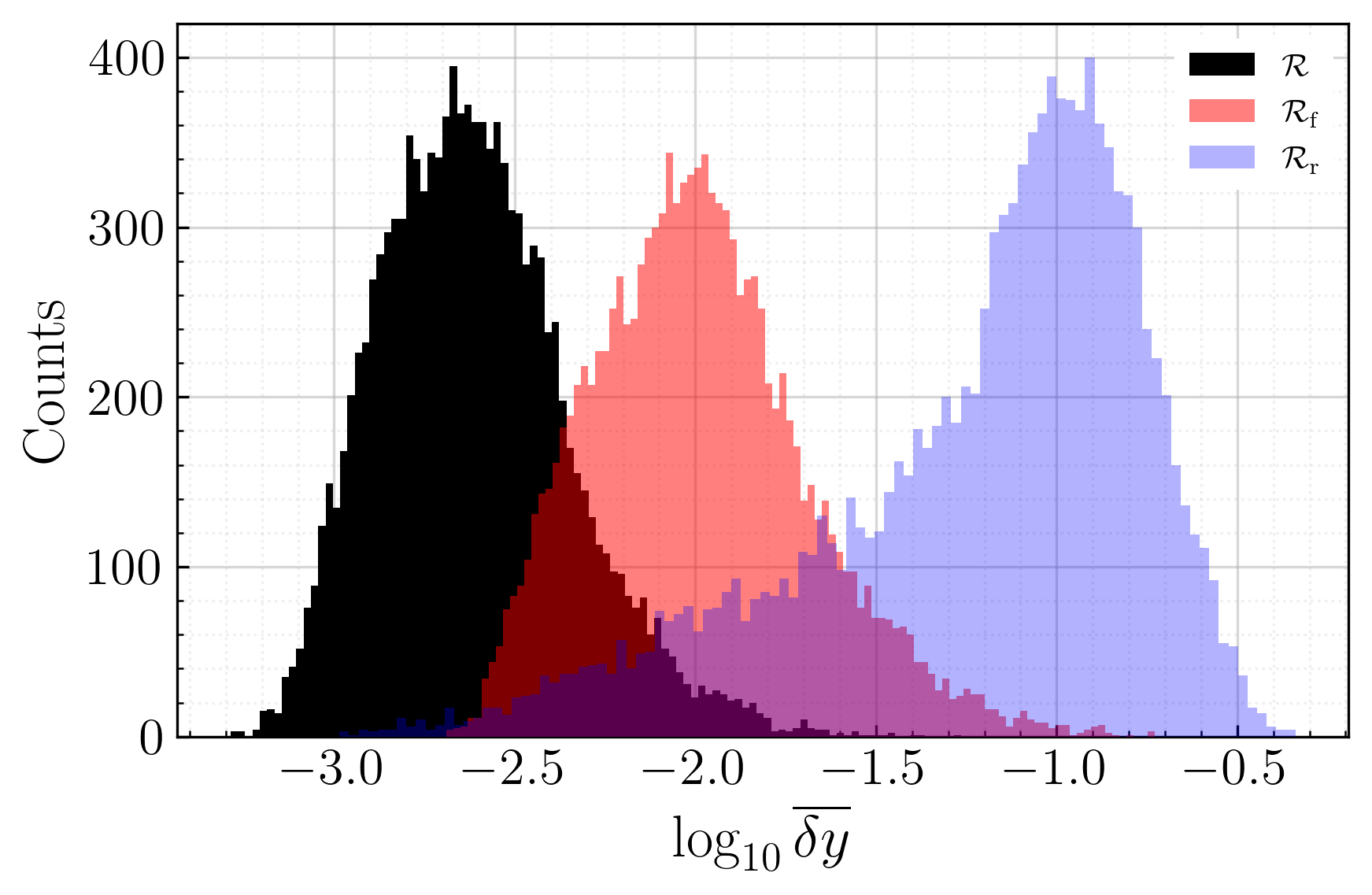}
\caption{\label{fig:anderson hist} Anderson testing set histogram for the MLP models
trained on the full $\mathcal{R},$ the FPS $\mathcal{R}_\mathrm{f},$ and
random-sampled $\mathcal{R}_\mathrm{r}$ training sets.}
\end{figure*}

\begin{figure*}[p] 
\centering
\includegraphics[width=\linewidth]{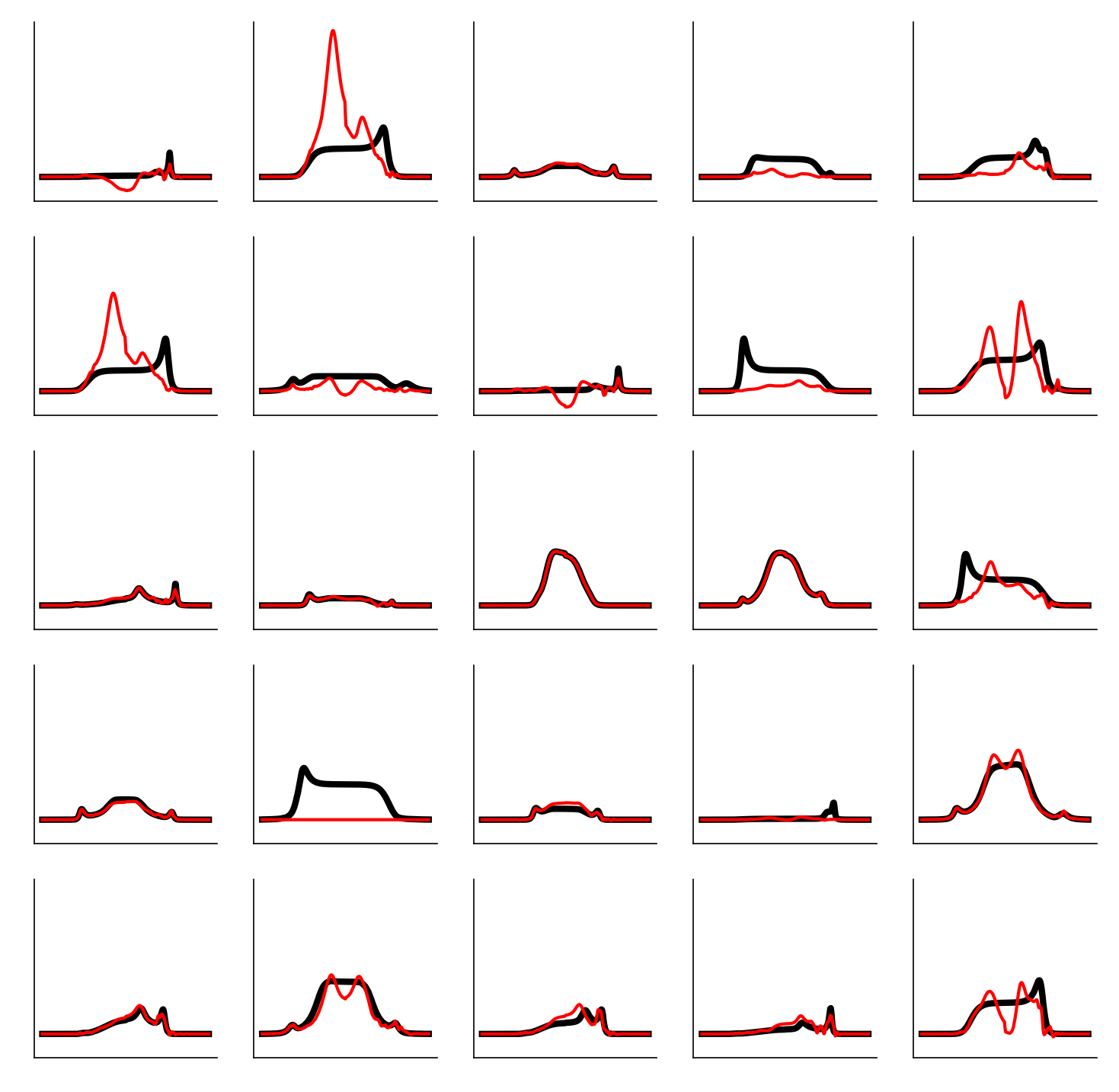}
\caption{\label{fig:anderson 25 full R krr} Random samples on the Anderson testing set for the KRR method trained on $\mathcal{R}_\mathrm{r}.$ Black is ground truth and red is prediction.}
\end{figure*}

\begin{figure*}[p] 
\centering
\includegraphics[width=\linewidth]{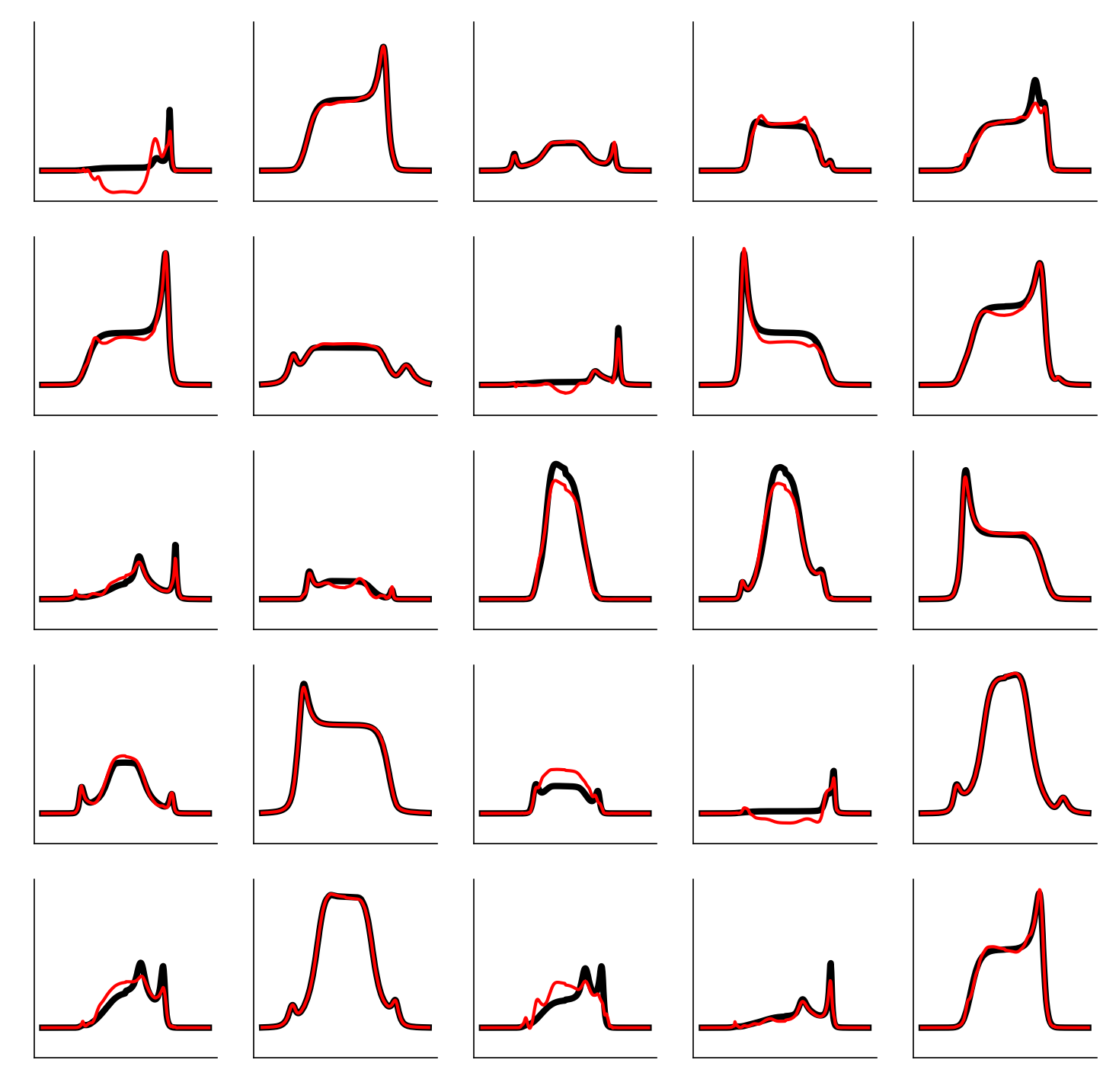}
\caption{\label{fig:anderson 25 full F krr} Random samples on the Anderson testing set for the KRR method trained on $\mathcal{R}_\mathrm{f}.$ Black is ground truth and red is prediction.}
\end{figure*}

\begin{figure*}[p] 
\centering
\includegraphics[width=\linewidth]{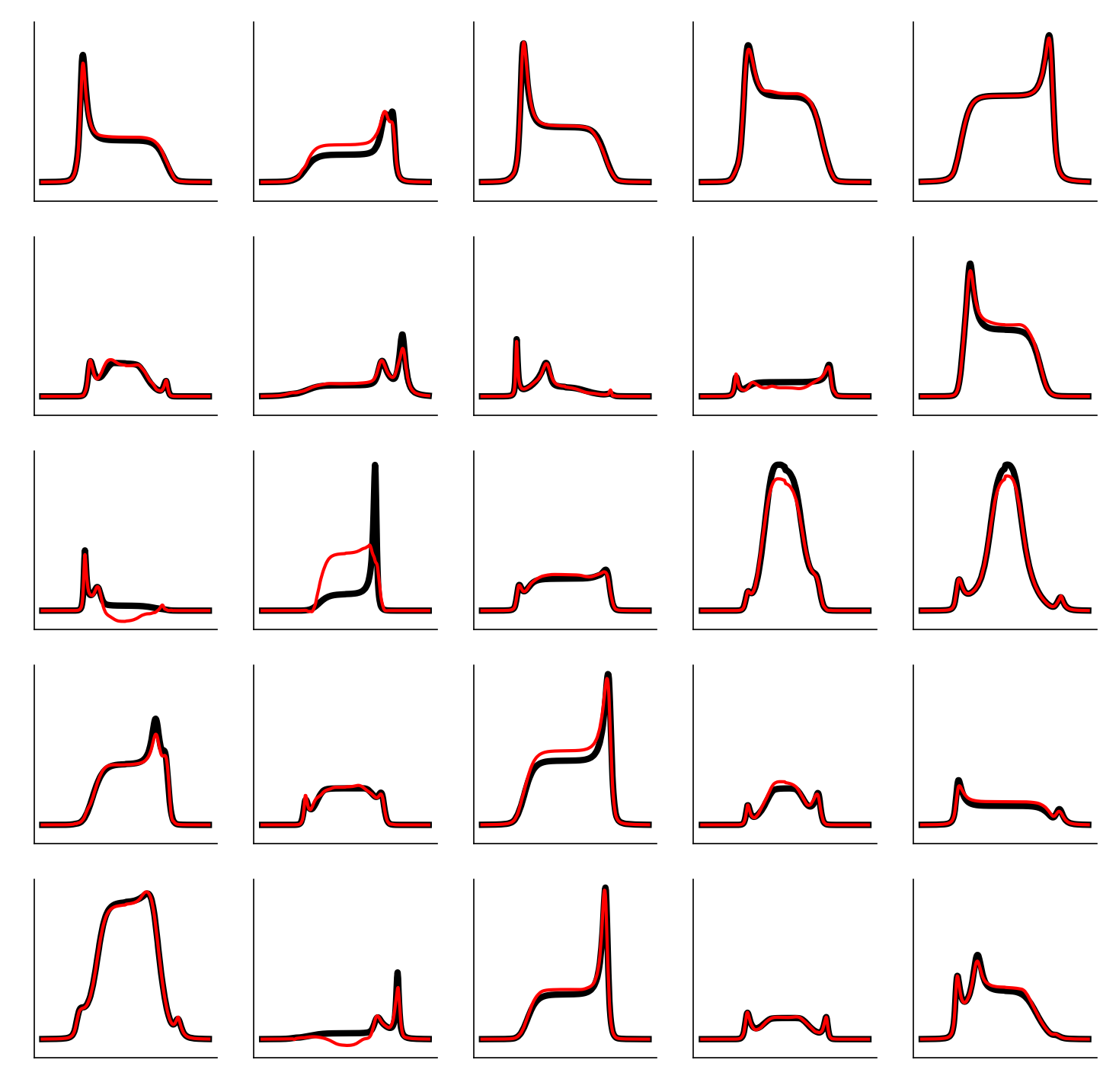}
\caption{\label{fig:anderson 25 full krr} Random samples on the Anderson testing set for the DC-KRR method trained on $\mathcal{R}_\mathrm{F}.$ Black is ground truth and red is prediction. Note that the full training set is used, but the data is chunked in the order of the FPS data points.}
\end{figure*}

\begin{figure*}[p] 
\centering
\includegraphics[width=\linewidth]{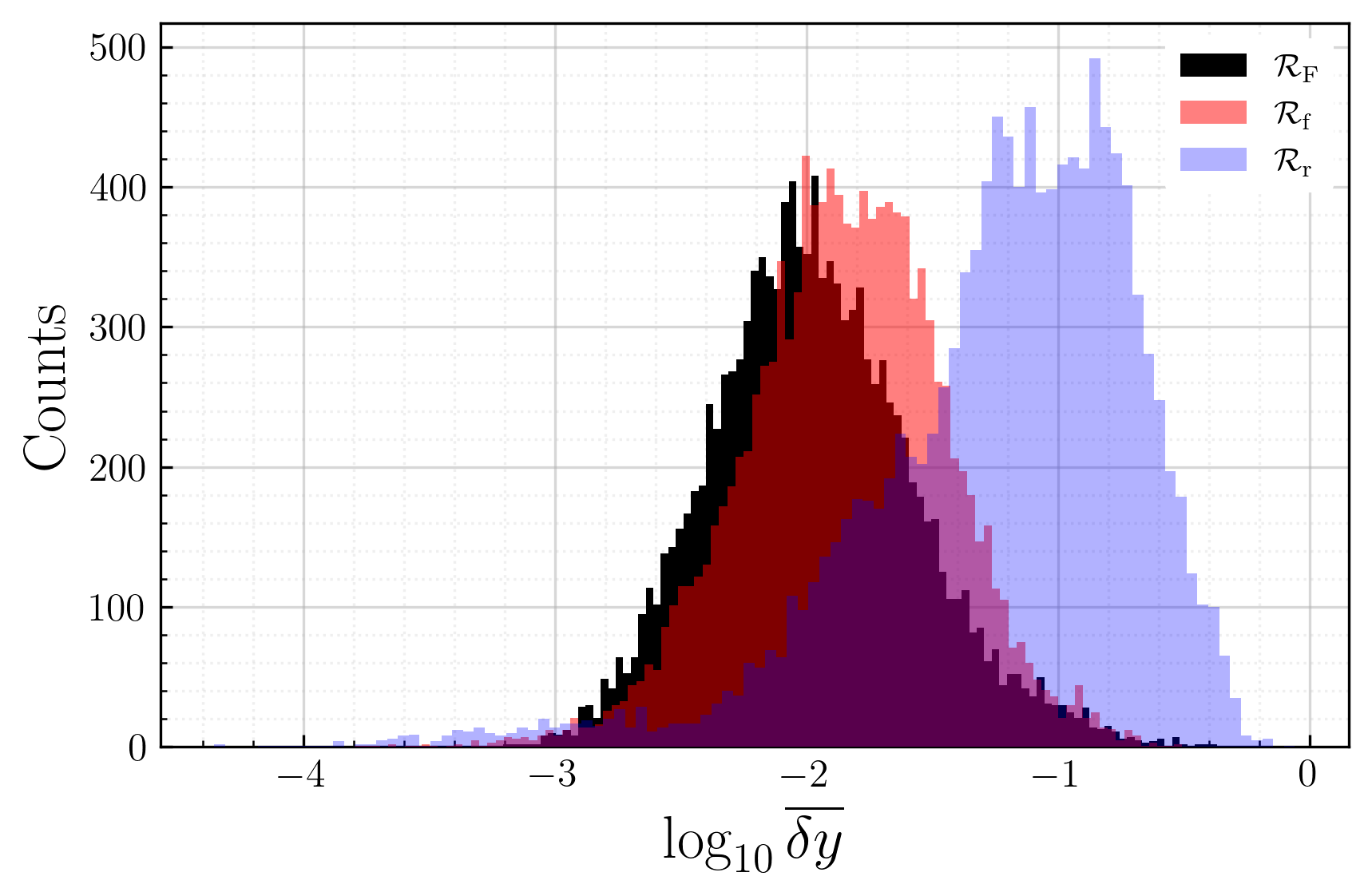}
\caption{\label{fig:anderson hist krr} 
Anderson testing set histogram fro the DC-KRR trained with the chunked FPS-ordered set full $\mathcal{R}_\mathrm{F}$ (black), and the KRR models trained on the random $\mathcal{R}_\mathrm{r}$ (red) and FPS ordered $\mathcal{R}_\mathrm{f}$ (blue) down-sampled sets. 
}
\end{figure*}

\begin{figure*}[p] 
\centering
\includegraphics[width=\linewidth]{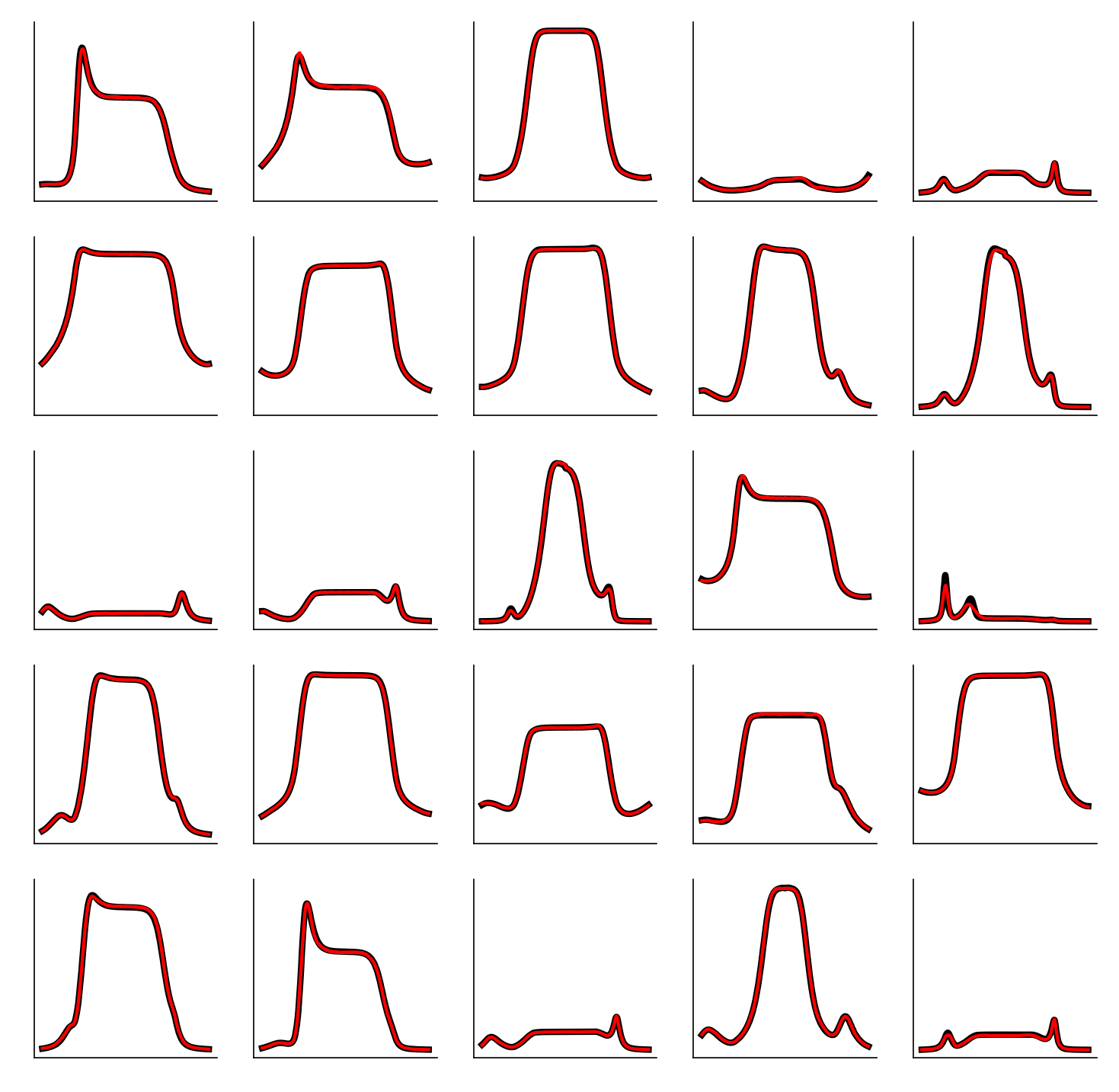}
\caption{\label{fig:kondo 25 full} Random samples on the Kondo testing set for an
MLP trained on $\mathcal{R}.$ Black is ground truth and red is prediction.}
\end{figure*}

\begin{figure*}[p] 
\centering
\includegraphics[width=\linewidth]{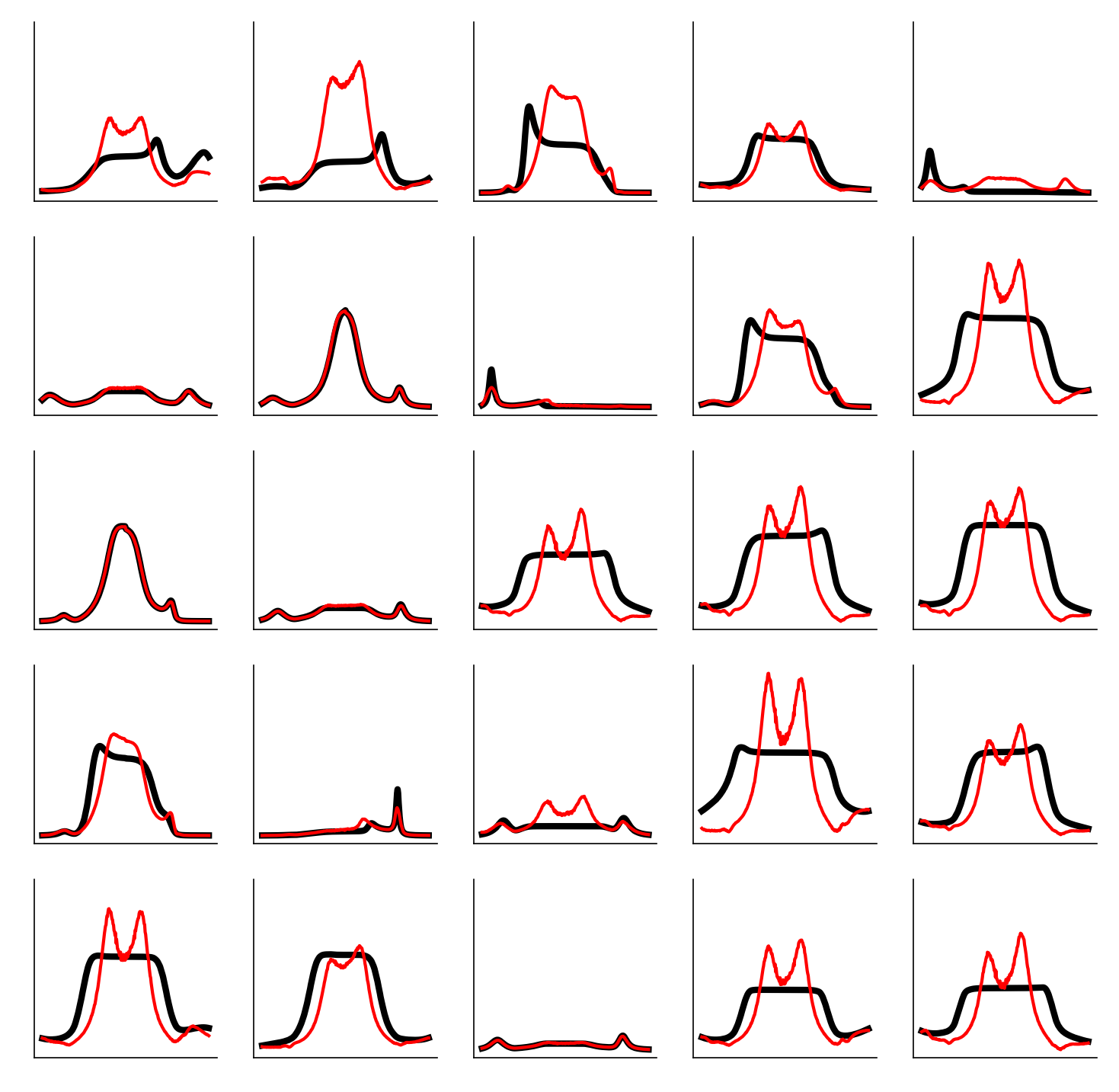}
\caption{\label{fig:kondo 25 full R} Random samples on the Kondo testing set for an
MLP trained on $\mathcal{R}_\mathrm{r}.$ Black is ground truth and red is prediction.}
\end{figure*}

\begin{figure*}[p] 
\centering
\includegraphics[width=\linewidth]{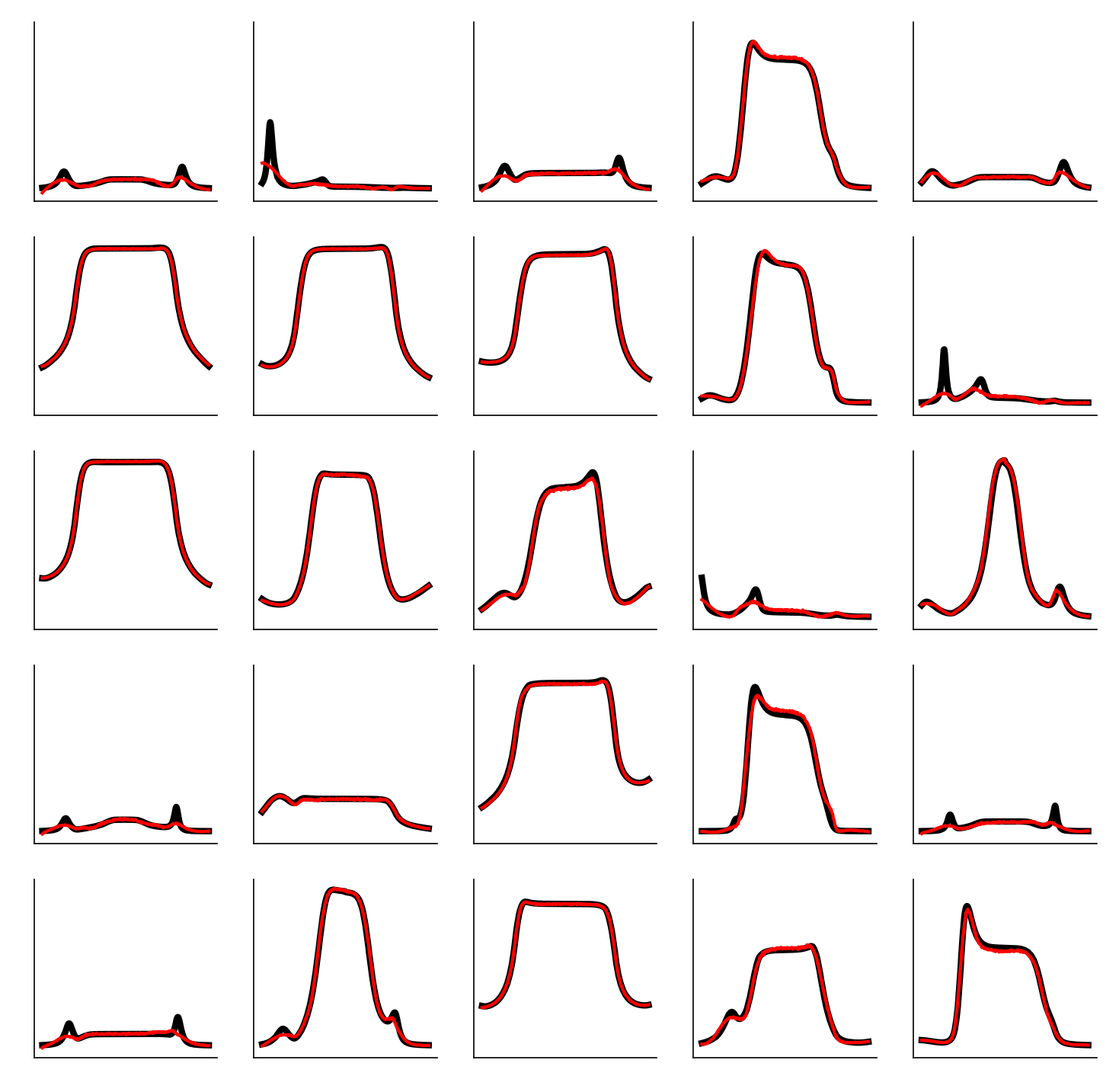}
\caption{\label{fig:kondo 25 full F} Random samples on the Kondo testing set for an
MLP trained on $\mathcal{R}_\mathrm{f}.$ Black is ground truth and red is prediction.}
\end{figure*}

\begin{figure*}[p] 
\centering
\includegraphics[width=\linewidth]{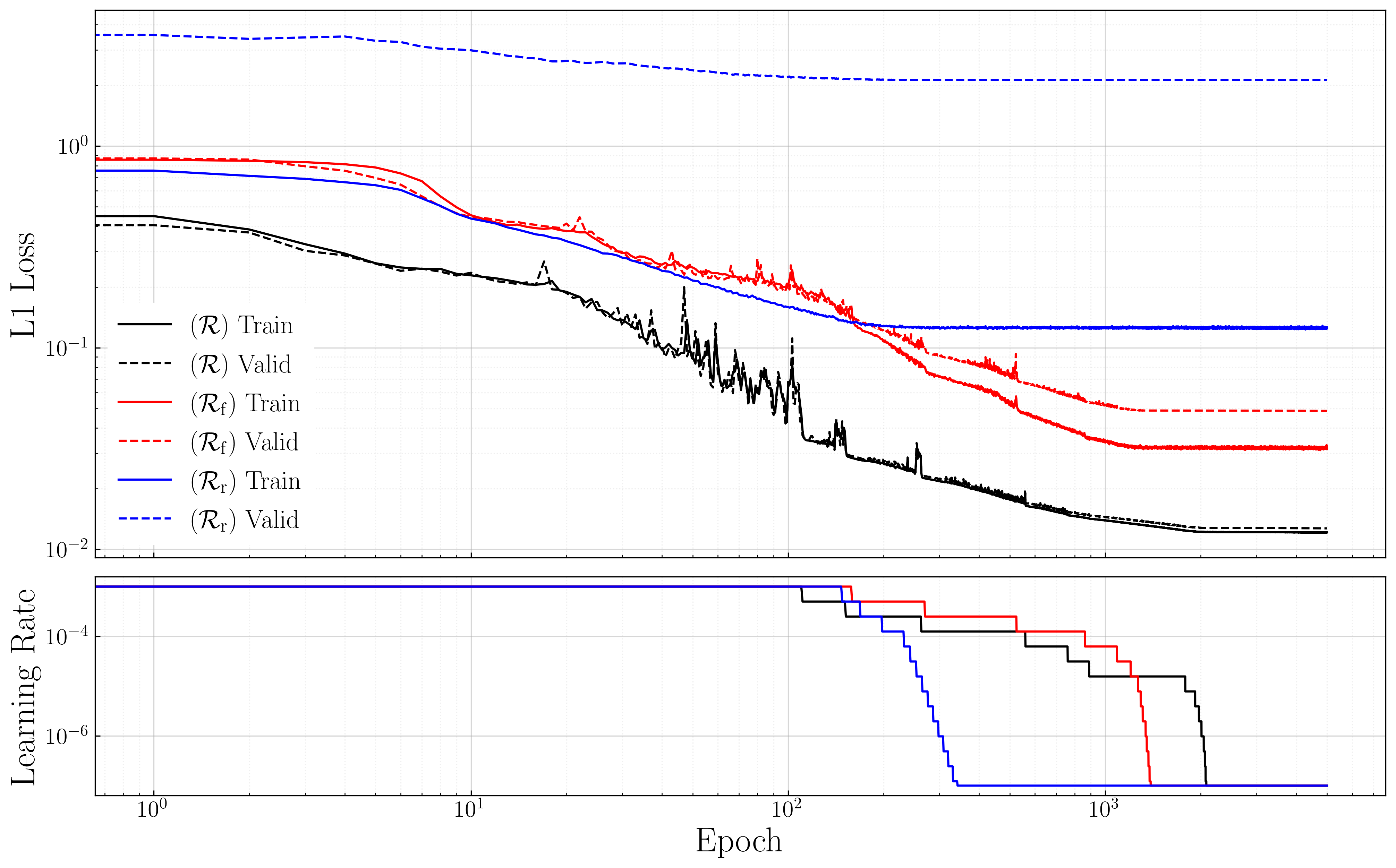}
\caption{\label{fig:kondo losses} Kondo training and validation losses and learning rates
plotted as a function of epochs for models trained on the full $\mathcal{R},$ FPS $\mathcal{R}_\mathrm{f},$ and random-sampled $\mathcal{R}_\mathrm{r}$ training sets.}
\end{figure*}

\begin{figure*}[p] 
\centering
\includegraphics[width=\linewidth]{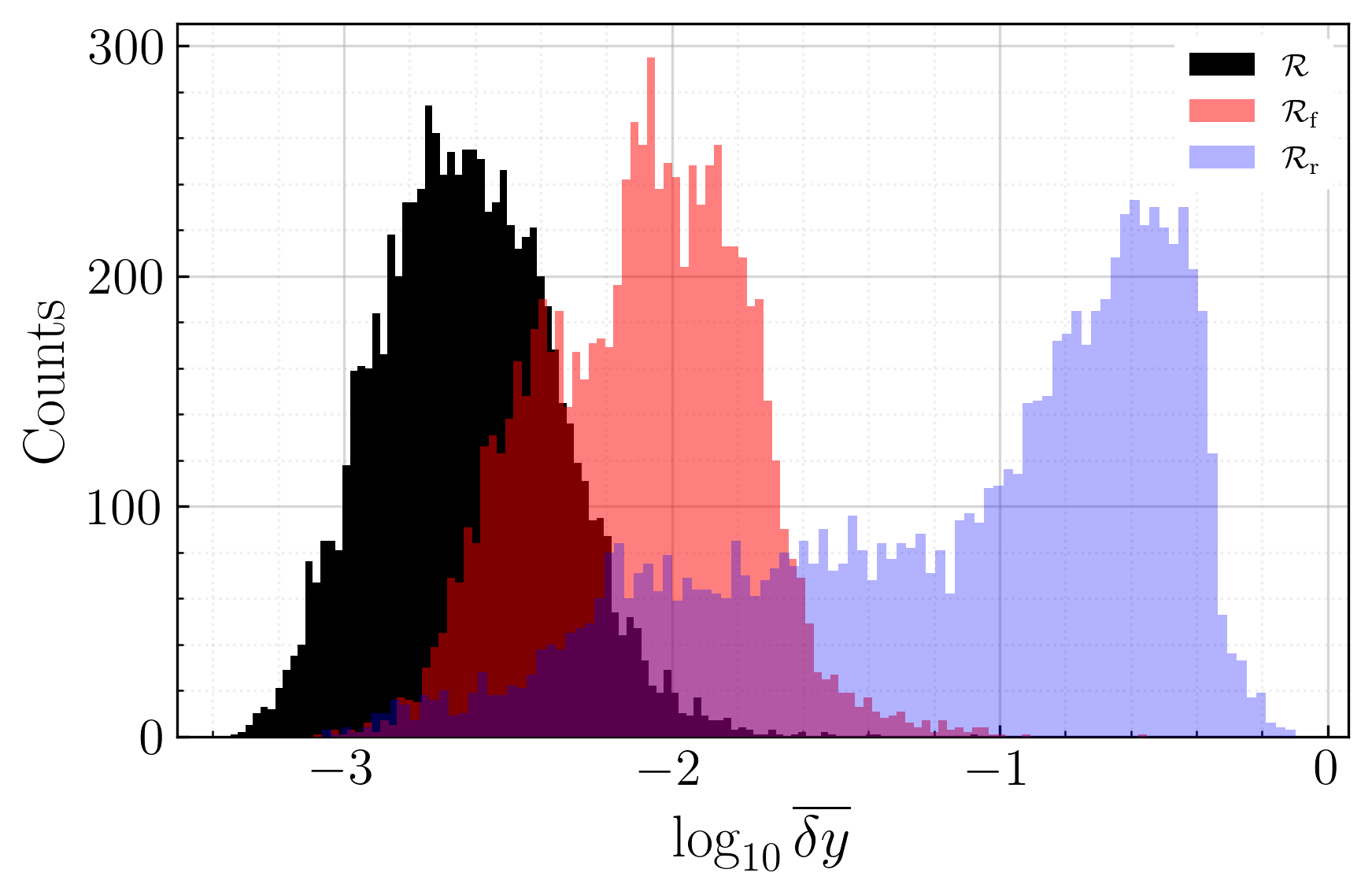}
\caption{\label{fig:kondo hist} Kondo testing set histogram for the MLP models
trained on the full $\mathcal{R},$ FPS $\mathcal{R}_\mathrm{f},$ and
random-sampled $\mathcal{R}_\mathrm{r},$ training sets.}
\end{figure*}

\begin{figure*}[p] 
\centering
\includegraphics[width=\linewidth]{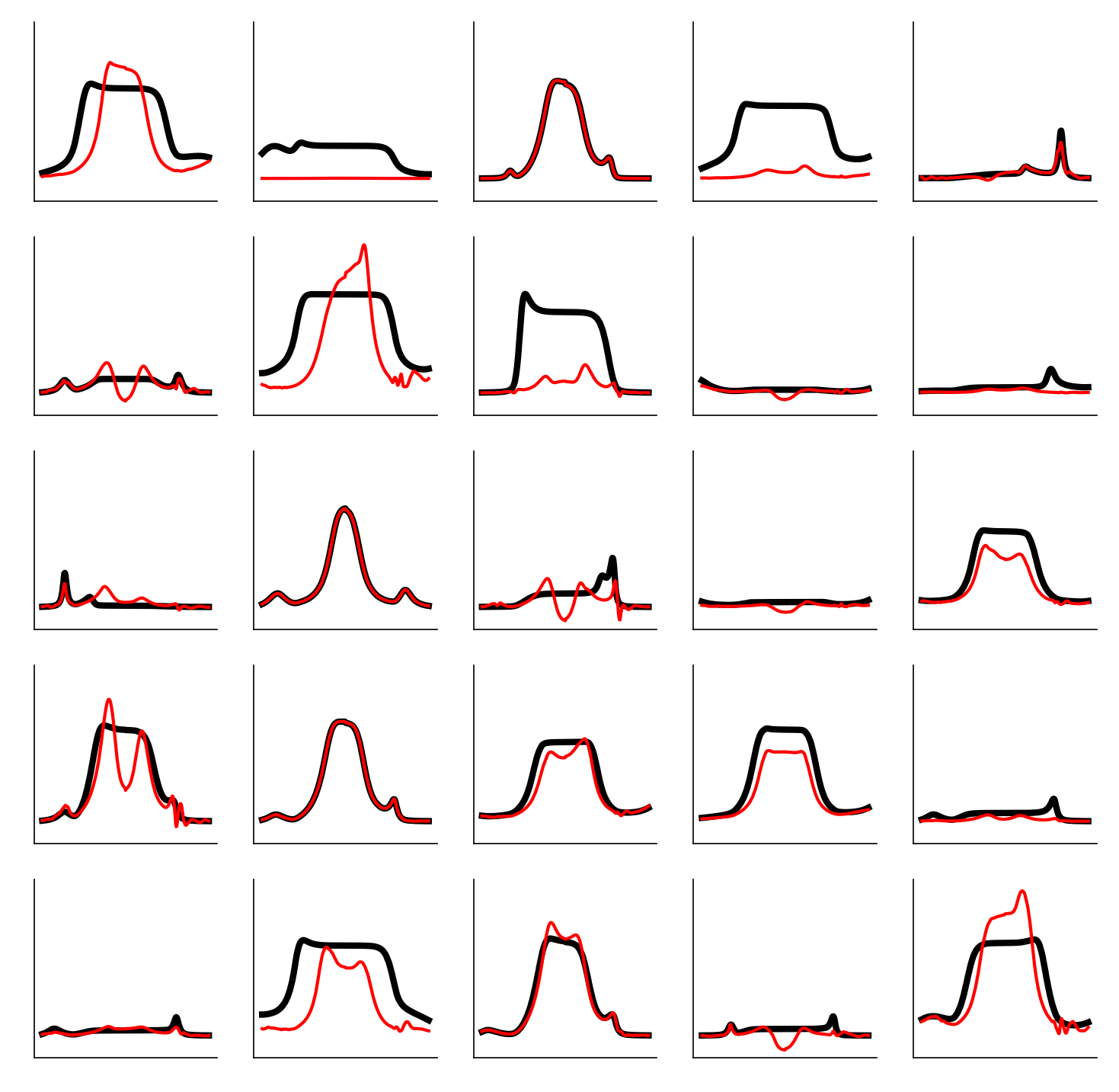}
\caption{\label{fig:kondo 25 full R krr} Random samples on the Kondo testing set for the KRR method trained on $\mathcal{R}_\mathrm{r}.$ Black is ground truth and red is prediction.}
\end{figure*}

\begin{figure*}[p] 
\centering
\includegraphics[width=\linewidth]{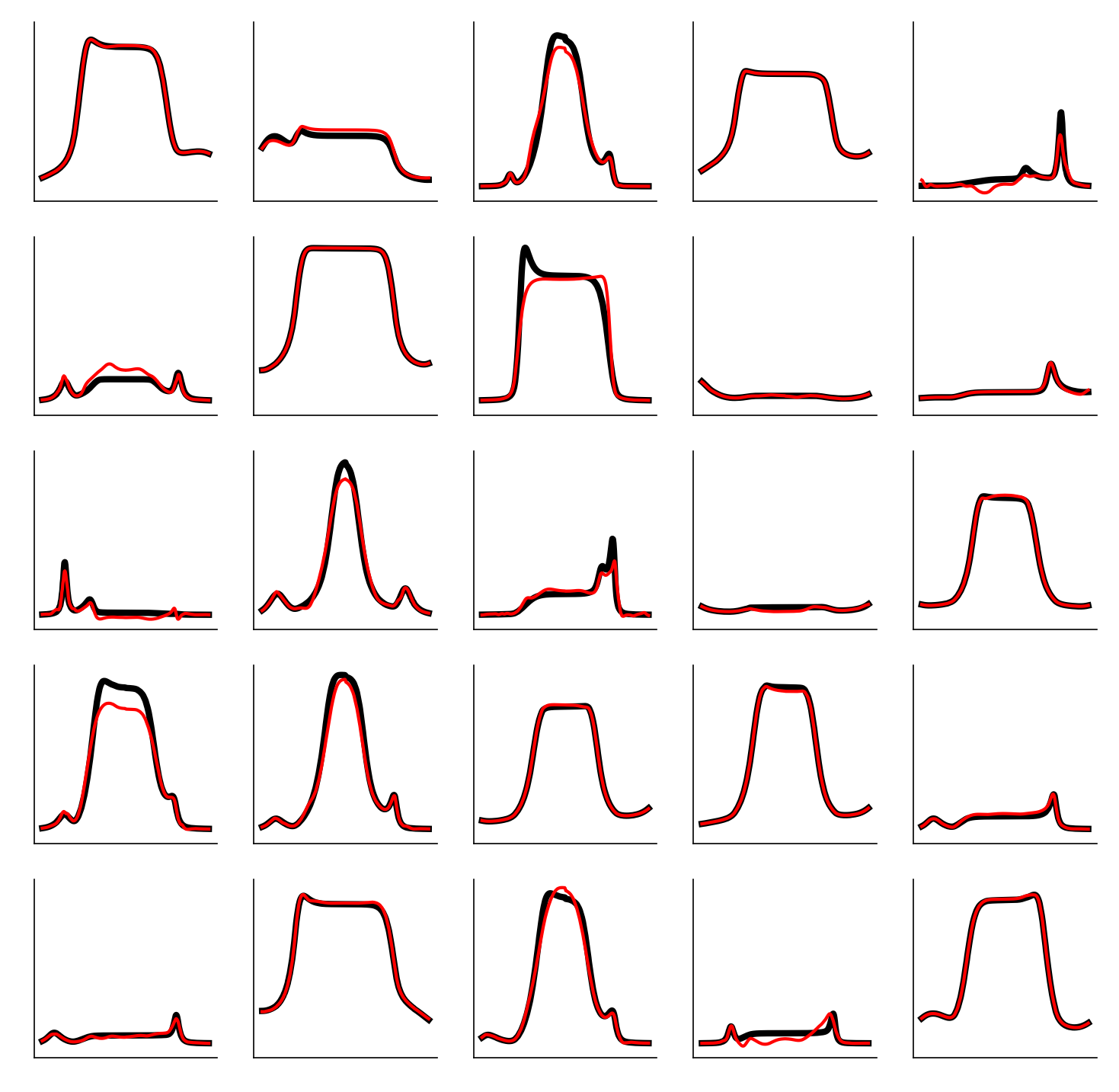}
\caption{\label{fig:kondo 25 full F krr} Random samples on the Kondo testing set for the KRR method trained on $\mathcal{R}_\mathrm{f}.$ Black is ground truth and red is prediction.}
\end{figure*}

\begin{figure*}[p] 
\centering
\includegraphics[width=\linewidth]{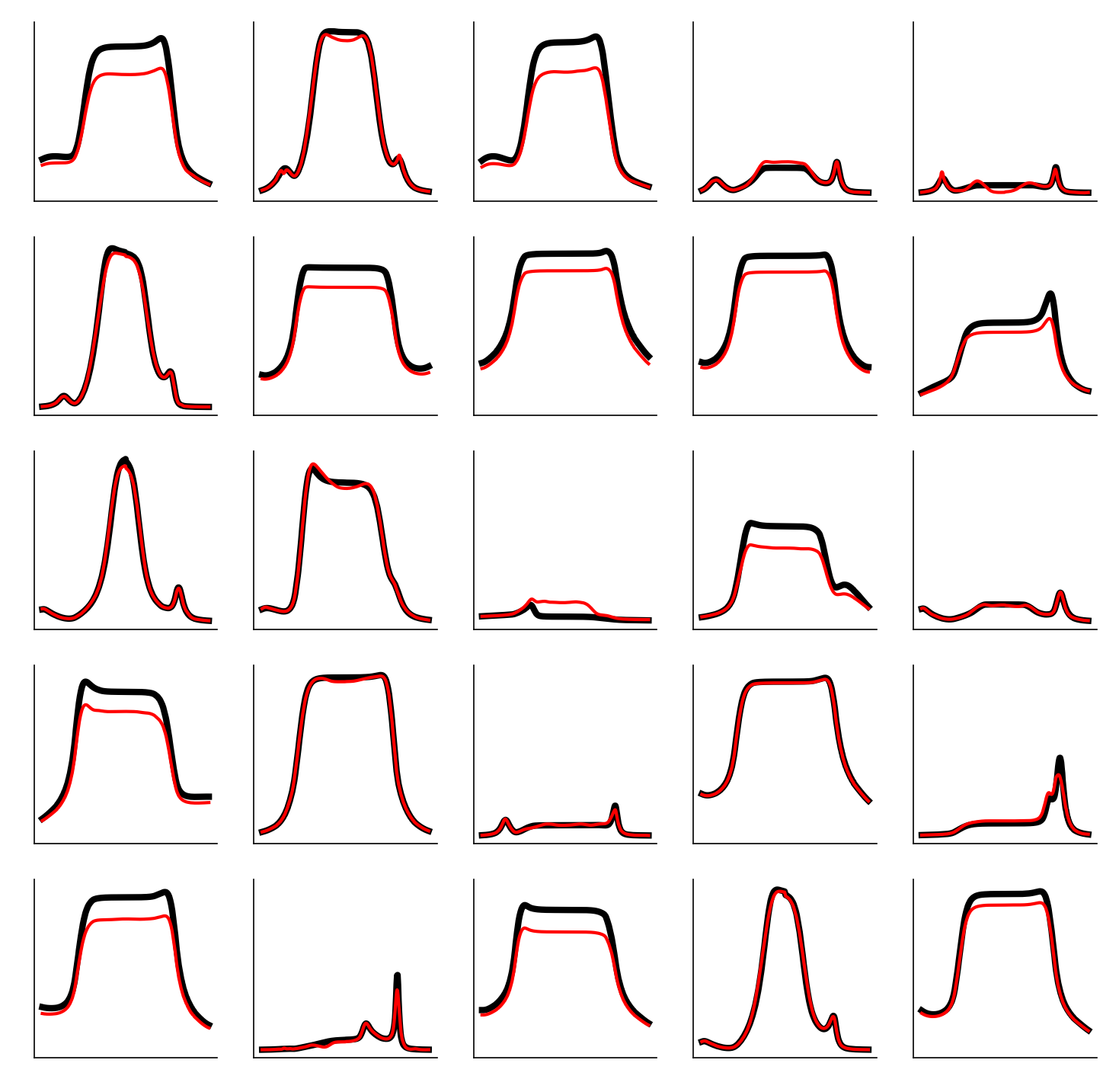}
\caption{\label{fig:kondo 25 full krr} Random samples on the Kondo testing set for the DC-KRR method
trained on $\mathcal{R}_\mathrm{F}.$ Black is ground truth and red is prediction. Note that the full training set is used, but
the data is chunked in the order of the FPS data points.}
\end{figure*}

\begin{figure*}[p]
\centering
\includegraphics[width=\linewidth]{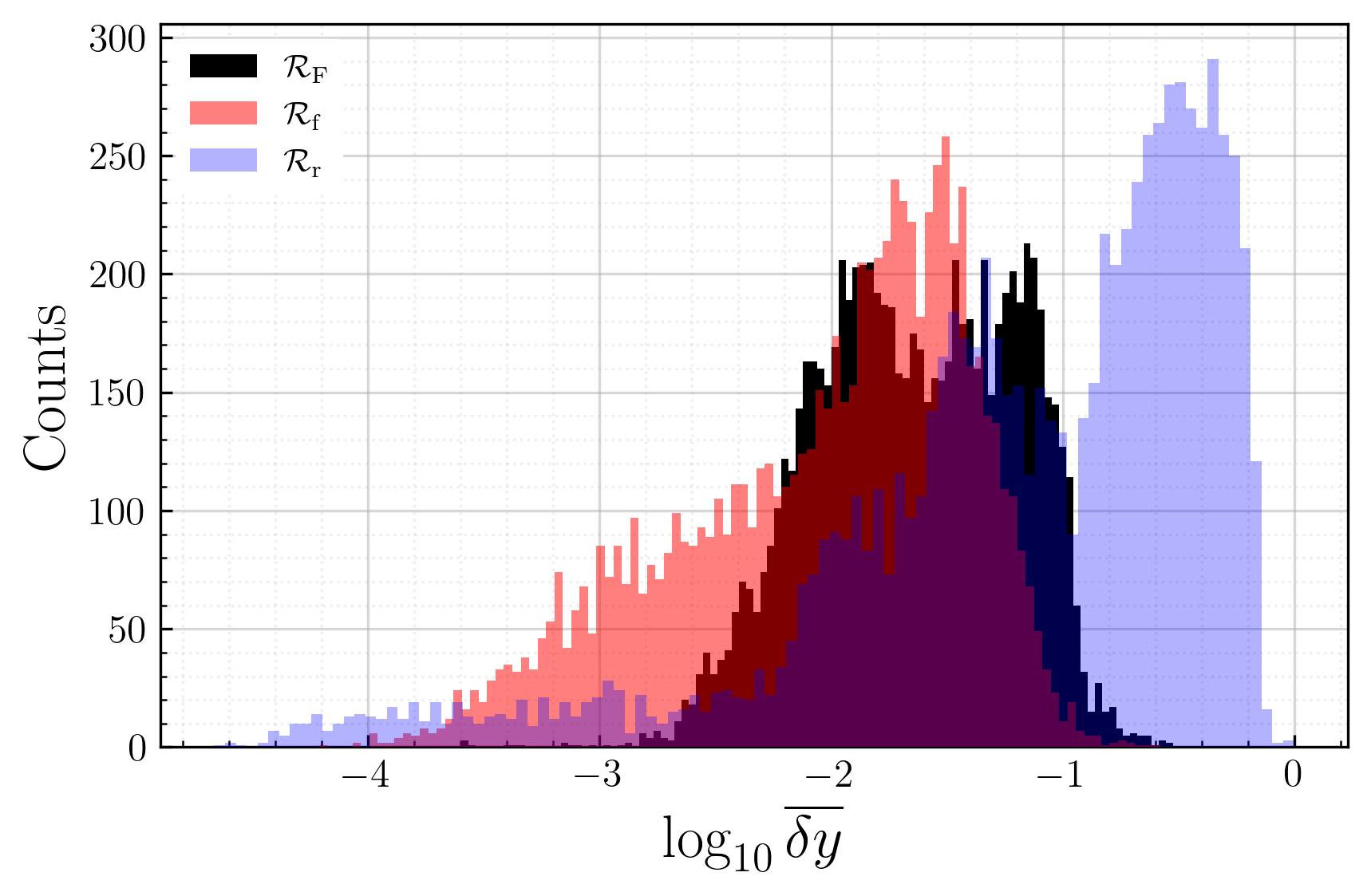}
\caption{\label{fig:kondo hist krr} 
Kondo testing set histogram fro the DC-KRR trained with the chunked FPS-ordered full set $\mathcal{R}_\mathrm{F}$ (black), and the KRR models trained on the random $\mathcal{R}_\mathrm{r}$ (red) and FPS ordered $\mathcal{R}_\mathrm{f}$ (blue) down-sampled sets. 
}
\end{figure*}

\end{appendix}

\bibliography{bib}

\end{document}